\documentclass[aps,preprint,nofootinbib]{revtex4-2}

\usepackage{graphicx}
\usepackage{dcolumn}
\usepackage{bm}

\usepackage[utf8]{inputenc}
\usepackage[T1]{fontenc}
\usepackage{mathptmx}
\usepackage{etoolbox}

\usepackage{caption}
\usepackage{epsfig,color}
\usepackage{multirow}
\usepackage{bbm}
\usepackage{enumerate}
\usepackage{verbatim}
\usepackage{braket}
\usepackage{mathtools,feynmp-auto,subcaption}
\usepackage{comment}
\usepackage{hyperref}
\hypersetup{colorlinks=true,linkcolor=blue,anchorcolor=blue,citecolor=blue,filecolor=blue,urlcolor=blue,bookmarksnumbered=true,pdfview=FitB}
\usepackage[english]{babel}
\usepackage[maxfloats=256]{morefloats}
\usepackage{lmodern}
\usepackage{mathptmx, newtxtext, newtxmath, xspace}
\usepackage[dvipsnames]{xcolor}
\usepackage[export]{adjustbox}
\newcommand{\ve}{\varepsilon}

\newcommand{\vf}{\varv_{\mathrm{F}}}
\newcommand{\kf}{k_{\mathrm{F}}}
\newcommand{\Ef}{E_{\mathrm{F}}}

\newcommand{\bk}{{\bf k}}

\newcommand{\bR}{{\bf R}}
\newcommand{\bG}{{\bf G}}
\newcommand{\bK}{{\bf K}}

\newcommand{\Wt}{W_{\bk_1,\bk_2;\bk'_1,\bk'_2}}

\newcommand{\nn}{\nonumber}
\newcommand{\beq}{\begin{equation}}
\newcommand{\eeq}{\end{equation}}
\newcommand{\bea}{\begin{eqnarray}}
\newcommand{\eea}{\end{eqnarray}}
\newcommand{\bse}{\begin{subequations}}
\newcommand{\ese}{\end{subequations}}
\newcommand{\bwt}{\begin{widetext}}
\newcommand{\ewt}{\end{widetext}}

\newcommand{\I}{\mathrm{Im}}
\newcommand{\R}{\mathrm{Re}}

\newcommand{\bsu}{\begin{subequations}}
\newcommand{\esu}{\end{subequations}}

\newcommand{\br}{{\bf r}}

\newcommand{\bi}{\begin{itemize}}
\newcommand{\ei}{\end{itemize}}

\renewcommand{\Re}{\operatorname{Re}}

\begin{document}

\title{Integrals of Products of Bessel Functions:\\ 
An Insight from the Physics of Bloch Electrons}
\author{Joshua Covey and Dmitrii L. Maslov}
\affiliation{Department of Physics, University of Florida, Gainesville, FL 32611-8440, USA}
\begin{abstract}
Integrals of products of Bessel functions exhibit an intriguing feature: 
under certain conditions on the parameters specifying the integrand,
they vanish identically. We provide a physical interpretation of this feature in the context of both single-particle and many-body properties of electrons on a lattice (``Bloch electrons''), namely, in terms of their density of states and umklapp scattering rate.  (In an  umklapp event, the change in the  momentum of two colliding electrons is equal to a reciprocal lattice vector, which gives rise to a finite resistivity due to electron-electron interaction.) In this context, the vanishing of an integral follows simply from the condition that either the density of states vanishes due to the electron energy lying outside the band in which free  propagation of electron waves is allowed, or that an umklapp process is kinematically forbidden due to the Fermi surface being smaller than a critical value.
\end{abstract}
\email{jcovey@ufl.edu}
\date{\today}

\raggedbottom

\maketitle 

\tableofcontents

\section{\label{sec:intro}Introduction} 
Bessel functions appear time and again in mathematics, physics, and engineering. The subject of this paper is a particular property of integrals of products of Bessel functions to vanish for 
a whole range of parameters. Various integrals of this kind were investigated via ingenious use of contour integral techniques by 
$19^\mathrm{th}$ century mathematicians, such as Hankel, Sonine, Weber, and Gegenbauer, to name a few, and later presented in Watson's seminal text on Bessel functions \cite{Watson1944AFunctions}. More examples, involving both cylindrical and spherical Bessel functions, were studied in more recent works \cite{Jackson1972IntegralsFunctions,Exton1979AIntegral,Mehrem1991AnalyticFunctions,Mehrem2010AFunctions,Mehrem2013AnalyticalFunctions,Majumdar2019WhenIntegrals}. 

\begin{figure}[t]
  \includegraphics[scale=.5]{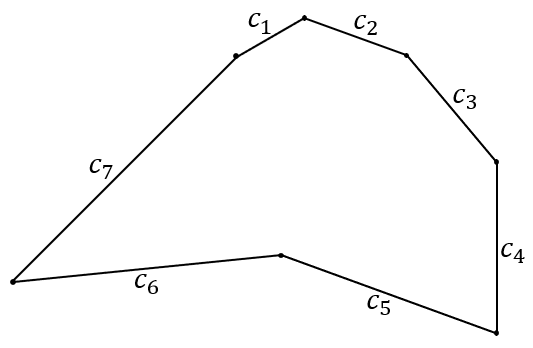}
  \caption{An illustration of the constraint \eqref{PC} for a seven-sided polygon  with sides satisfying $c_1<c_2\dots c_6<c_7$. 
  }
\label{fig:tri constraint}
\end{figure}

The most general form of such integrals for cylindrical Bessel functions of the first kind may be written as
\begin{equation}
    \begin{split}
   f(\{c_n\})=
\int^\infty_0 d\rho \rho^{\alpha-1} \prod_{n=1}^{M>1}J_{\nu_n}(c_n \rho), \label{alpha}     
    \end{split}
\end{equation}
where 
the coefficients $c_n$ are real and positive, and ordered such that $c_1\leq c_2\leq \dots\leq c_{M-1}<c_M$, while 
the ``dimensionality'' $\alpha$ and indices $\nu_n$ may take arbitrary, including complex, values.   (If $\alpha$ is real, it can indeed be viewed as the dimensionality of some abstract geometrical space, but we will be using the same term even for an arbitrary $\alpha$.)
The integral \eqref{alpha} is 
convergent 
if $-\sum_{n=1}^{M}\Re\nu_n<\Re\alpha<M/2+1$. 
The result for the integral depends crucially on two relations 
between the input parameters. The first one, which we will be referring to as the "polygonal constraint", 
reads
\begin{equation}
    \begin{gathered}
       c_M<\sum_{n=1}^{M-1} c_n,
        \label{PC}
    \end{gathered}
\end{equation}
in which case the coefficients $c_n$ can be viewed as the sides of a planar polygon, see Fig.~\ref{fig:tri constraint}. 
The second one, which we
will be referring to as the "charge neutrality condition",
is given by
\bea
\nu_M=\sum_{n=1}^{M-1}\nu_n \ + \ \alpha \ - \ 2m; \ \ m=1,2,3\dots
 \label{CNC}
 \eea
 If we assign a fictitious "charge" $\nu_n$ to the $
 n^{\mathrm{th}}$ side of the polygon, the condition \eqref{CNC} says that the charge on the $M^{\mathrm{th}}$ side must be equal to the sum of the charges on other sides, modulo the integration dimensionality and a negative, even integer. 
 
A closed-form solution of Eq.~\eqref{alpha} for the case when the polygonal constraint \eqref{PC} is \emph{not} satisfied was 
apparently derived first by Exton and Srivastava \cite{Exton1979AIntegral}, and can also be
found in the monumental tables
by Prudnikov, Brychkov, and Marichev  \cite{
prudnikov-integrals-table}:
\begin{flalign}
\begin{split}
f(\{c_n\})&=\begin{cases} 
2^{\alpha-1}c_M^{\nu_M-\mu}\frac{\Gamma(\mu/2)}{\Gamma\left(\nu_M-\mu/2+1\right)}\prod_{n=1}^{M-1}\frac{c_n^{\nu_n}}{\Gamma(\nu_n+1)} \\
\hspace{0.08\textwidth} \times \ F_C^{(M-1)}\left(\frac{\mu}{2},\frac{\mu}{2}-\nu_M;\nu_1+1,\ldots,\nu_{M-1}+1;\frac{c_1^2}{c_M^2},\ldots,\frac{c_{M-1}^2}{c_M^2}\right),\;\mathrm{if}\;\eqref{CNC}\;\mathrm{is\;false}; \\
0 \hspace{0.402\textwidth},\;\mathrm{if}\;\eqref{CNC}\;\mathrm{is\;true},
\end{cases}
\label{alpha1}
\end{split}
\end{flalign}
where $\mu=\alpha+\sum_{n=1}^M\nu_n$ and  
$F_C^{(N)}(a,b;x_1,\dots, x_N;y_1,\dots, y_N)$ is the "type $C$" Lauricella hypergeometric function. The null result in the second line of Eq.~\eqref{alpha1} follows from the first line, for when Eq.~\eqref{CNC} is satisfied the gamma function $\Gamma(\nu_M-\mu/2+1)=\Gamma(-m+1)$ has a pole.
If the polygonal constraint \emph{is} satisfied, 
there is no general closed-form of the integral \eqref{alpha}, except for $M=2,3,4$ and then for some special choices of $\nu_n$ and $\alpha$,
considered by Sonine \cite{Sonine1880RecherchesSeries}, Nicholson \cite{Nicholson1920GeneralisationSonine}, and Watson \cite{Watson1944AFunctions}.

The peculiar, yet sublime, emergence of the geometric constraint \eqref{PC} in tandem with the 
charge-neutrality condition \eqref{CNC} appears in a variety of seemingly different physical problems
for the particular case of integrals over products of integer- or half-integer-order Bessel functions. It is in these cases that the integrand can be expressed in terms of $\delta$-functions,
a tool $19^{\text{th}}$ century mathematicians did not have at their disposal. Integrals of the same form as \eqref{alpha} with $\nu_n\in \mathbb{Z}$ and $\nu_n+1/2\in\mathbb{Z}$ arise in a variety of settings. In particular, they have been analyzed by Jackson and Maximon \cite{Jackson1972IntegralsFunctions} with the help of $\delta$-functions for the special case of $M=3$ and either $\nu_n\in\mathbb{Z}$ in two dimensions or $\nu_n + 1/2 \in\mathbb{Z}$ in three dimensions. Mehrem \cite{Mehrem1991AnalyticFunctions,Mehrem2010AFunctions,Mehrem2013AnalyticalFunctions} considered analytical solutions to integrals of products of up to four spherical Bessel functions,   arising in the context of nuclear scattering. 
Integrals of the same form as \eqref{alpha} may also be found in other unsuspected places, such as in the random-walk problem \cite{Exton1979AIntegral,Majumdar2019WhenIntegrals}.

A similar class of integrals involves a product of trigonometric and Bessel functions, e.g.,
\bea
g(\{c_n\})=\int^\infty_0 d\rho \rho^{\alpha-1}\left[A\cos (c_M\rho)+B\sin (c_M\rho)\right] \Pi_{n=1}^{M-1} J_{\nu_n}(c_n\rho).\label{g}
\eea
Integrals of this type were studied in the random-walk problem
\cite{Majumdar2019WhenIntegrals}.
To the best of our knowledge, explicit results are known only for $M=2,3$   \cite{prudnikov-integrals-table}. In particular, for $B=0$, $M=3$, $\alpha=1$ and $\nu_1=\nu_2$, the integral is non-zero only if the polygonal constraint \eqref{PC} is satisfied, i.e., $c_3<c_1+c_2$, in which case the result is expressed via the complete elliptic integrals.

The vanishing of the integrals \eqref{alpha} and \eqref{g} for a whole range of parameters is an interesting phenomenon, particularly because it does not follow from any orthogonality condition. 
The goal of this paper is to give a physical interpretation of this phenomenon in the context of
properties 
of electrons on a lattice ("Bloch electrons"). 

In the absence of interaction, 
the electron wavefunction satisfies the Bloch theorem \cite{AshcroftMermin}
\beq
\psi_{n\bk}(\br+\bR)=e^{i\bk\cdot\bR}\psi_{n\bk}(\br),
\eeq
where $\bR$ is any lattice vector and the quasimomentum $\bk$ is defined up to an arbitrary reciprocal lattice vector $\bG$, the latter satisfying the condition $e^{i\bG\cdot\bR}=1$. The unit cells of the reciprocal lattice are known as Brillouin zones and, conventionally, one considers physically nonequivalent quasimomenta within the first Brillouin zone. An electron can propagate through the lattice only if its energy 
lies within one of the bands, 
which are 
separated by forbidden energy gaps.
An all important property of the electron spectrum is the density of states
\bea
\nu(E)=\int\frac{d^Dk}{(2\pi)^D}\delta\left(E-\ve_{\bk}\right),\label{DOS}
\eea
where $\ve_{\bk}$ is the electron dispersion within a given band.
Naturally, $\nu(E)$ is non-zero only if $E$ is within the band, and zero otherwise. Already this simple property allows one to see a physical example of the polygonal constraint for integrals of the type  \eqref{g}.

We illustrate our approach for the simplest lattice model--the tight-binding model--in which electron hopping is allowed only between a limited number of nearest neighbors. For example, the energy spectrum of electrons for a square lattice with nearest-neighbor hopping is given by  \cite{Martin:book}
\bea
\ve_\bk=-t\left(\cos k_x+\cos k_y\right),\;-\pi\leq k_x,k_y\leq \pi,
\label{tb2D}
\eea
where we have set the lattice constant to unity, $t$ is the "hopping amplitude", and the first Brillouin zone is a square of side $\pi$. The band spans the interval of energies from $-2t$ to $2t$, such that the density of states is non-zero only if $-2t<E<2t$. To see how it is relevant to the integral \eqref{g}, we replace the $\delta$-functions in Eq.~\eqref{DOS} by its integral representation and swap the order of integration:
\bea
\nu(E)=\int^\infty_{-\infty} d\rho e^{iE\rho} \int \frac{dk_x}{2\pi} e^{i\rho t\cos k_x}\int \frac{dk_y}{2\pi} e^{i\rho t\cos k_y}.
\eea
Using the integral representation of a Bessel function of integer order
\beq
J_n(z) = \frac{1}{2\pi \ i^n}\int_0^{2\pi} d\phi e^{i(z\cos{\phi}\pm n\phi)}, 
\label{iota2}
\eeq
we obtain
\bea
\nu(E)=2\int^\infty_{0} d\rho \cos(E\rho) J_0^2(t\rho)=\frac{2}{\pi t}\Theta(2t-|E|)\mathbb{K}\left(\sqrt{1-\left(\frac{E}{2t}\right)^2}\right),\label{DOS2}
\eea
which is the special case of Eq.~\eqref{g} with $B=0$, $\alpha=1$, $M=3$, $c_3=E$, $\nu_1=\nu_2=0$, and $c_1=c_2=t$. Here, $\mathbb{K}(x)$ is the complete elliptic integral of the first kind. Now we see that the polygonal constraint is nothing but the condition for the energy to lie within the band. (Although we chose all $c_n>0$, the integral \eqref{DOS2} is obviously an even function of $E$, such that the interval $-2t<E<0$ does not require a special consideration.) In what follows, we show that the integral \eqref{g} vanishes for any $M$, $\alpha=1$, and under certain conditions on $\nu_n$'s, by considering the density of states and  conductivity of electrons on a hyperrectangular
lattice in $D=M-1$ dimensions.

The integral \eqref{alpha} occurs in a more complex physical context of mutual scattering of Bloch electrons. In contrast to the real momentum, which is conserved in electron-electron scattering in the absence of a lattice, the quasimomentum is conserved only up 
to an arbitrary reciprocal lattice vector. The relation between initial momenta $\bk_1\dots \bk_N$ and final momenta $\bk'_1\dots \bk'_N$ in an $N$-electron collision can be written as
\bea
\sum_{n=1}^N\left(\bk_n'-\bk_n\right)=\bG.\label{umklapp}
\eea
If $\bG=0$, the scattering process is called "normal". If $\bG\neq 0$, it is called "umklapp" (from German "umklappen", which means "fold down" in English). In the latter 
case, at least one of the final momenta lies outside of the first Brillouin zone, but it can be translated back ("folded fown'') to this zone via a shift by $\bG$. 
A very important distinction between normal and umklapp scattering processes is that only the latter can render the electrical resistivity of a 
metal finite, even in the absence of electron scattering by disorder and phonons \cite{physkin}. Due to the Pauli exclusion principle, scattering affects only electrons with energies in a narrow window near the Fermi energy ($\Ef$), of width set by the thermal energy, $T$. (Hereinafter, we set $k_B=1$.) For a two-electron collision, this implies that the scattering rate, and thus the resistivity, are proportional to $T^2$. 

Since the quasimomenta of electrons are confined to the Fermi surface, the condition \eqref{umklapp} for an umklapp scattering process can be satisfied only if the Fermi surface is sufficiently large. Namely, the umklapp scattering rate is equal to zero if
\bea
\left\vert\sum_{n=1}^N\left(\bk_n'-\bk_n\right)\right\vert_{\bk_1\dots\bk_N,\bk_1'\dots\bk'_N\in\mathrm{FS}}< G_0,\label{umklapp2}
\eea
where $G_0$ is the magnitude of the shortest non-zero reciprocal lattice vector, which is on the order of the inverse interatomic distance. The condition \eqref{umklapp2} is reminiscent of the polygonal constraint \eqref{PC}, and the vanishing of the umklapp scattering rate is reminiscent of the analogous property of the integral \eqref{alpha}. This is the basis of our physical interpretation.  

The rest of the paper is organized as follows.
In Sec.~\ref{sec:hyper}, we consider the density of states and conductivity of electrons on a $D$-dimensional lattice, thereby proving the polygonal constraint for a certain class of the integrals of type \eqref{g}.
In Sec.~\ref{ref:phys motiv}, we show that integrals of the form 
\eqref{alpha} do appear in the umklapp scattering rate with $M=2N+1$, integer or half-integer 
$\nu_n$, and $\alpha=2$ or $\alpha=3-M/2$ for 2D and 3D lattices, respectively, if the actual Fermi surfaces are replaced by circles in 2D and spheres in 3D.
In Secs.~\ref{ref:subsec two particle} and \ref{sec:N=2}, we consider a particular example of two-electron umklapp scattering on a 2D honeycomb lattice, encountered, e.g., in graphene \cite{CastroNeto2009TheGraphene}. $N$-electron scattering processes on 2D and 3D lattices are analyzed in Secs.~\ref{ref:subsec gen higher order} and \ref{ref:scatt 3d}, respectively. The threshold behavior of the scattering rate just above the threshold for umklapp scattering is derived in Sec.~\ref{ref:subsec thresh behav}. In Sec.~\ref{ref: math conseq}, we consider more mathematical aspects of the charge-neutrality condition \eqref{CNC}.
A detailed derivation of the umklapp scattering rate and further mathematical details are delegated to the supplementary material.

\section{\label{sec:hyper}Single-particle properties of Bloch electrons on a $D$-dimensional lattice}
In this section we analyze 
Eq.~\eqref{g}, using some properties of Bloch electrons on a hyperrectangular
lattice in $D=M-1$ dimensions. 
Such a lattice is formed by $D$ orthogonal vectors of length $a_n$, 
and we assume that electrons can hop only between nearest neighbors with amplitude $t_n$ in the $n^{\mathrm{th}}$ direction. Analogous to the 2D case [cf. Eq.~\eqref{tb2D}], the electron energy spectrum is given by
\bea
\ve_\bk=-\sum_{n=1}^{M-1}t_n \cos(k_n a_n),
\eea
and the band spans an energy interval $\left(-\sum_{n=1}^{M-1}t_n,\sum_{n=1}^{M-1}t_n\right)$.
The corresponding Brillouin zone is also a hyperrectangle
with sides $2\pi/a_n$. Following the same steps that lead us to Eq.~\eqref{DOS2}, we obtain for the density of states
\bea
\nu(E)&=&\Pi_{n=1}^{M-1} \int^{\pi/a_n}_{-\pi/a_n} \frac{dk_n}{2\pi}\delta\left(E-\sum_{n=1}^{M-1}t_n \cos(k_n a_n)\right)\nn\\
&=&2\Pi_{n=1}^{M-1} a_n^{-1}\int^\infty_0 d\rho \cos(E\rho)\Pi_{n=1}^{M-1} J_0(t_n\rho),\label{DOSHC}
\eea
which is non-zero only if $|E|<\sum_{n=1}^{M-1} t_n$. Therefore, we have proven that the integral \eqref{g} is non-zero only if Eq.~\eqref{PC} is satisfied for  arbitrary $M$,  provided that $B=0$ and $\nu_1=\dots=\nu_{M-1}=0$. 

In the context of Bloch electrons, the Bessel functions of non-zero (but integer) order occur in calculating averages of the projections of momenta or velocities. For example, the conductivity in the presence of short-range disorder, characterized by the mean free time $\tau_d$, is given by \cite{AshcroftMermin}
\bea
\sigma_{ij}(\Ef)={2e^2\tau_d}\Pi_{n=1}^{M-1} \int^{\pi/a_n}_{-\pi/a_n} \frac{dk_n}{2\pi}\delta\left(\Ef-\sum_{n=1}^{M-1}t_n \cos(k_n a_n)\right)\varv_i\varv_j,\label{cond}
\eea
where $\varv_
n=\partial \ve_\bk/\partial k_n=t_na_n\sin(k_na_n)$ is the $n^\mathrm{th}$ Cartesian component of the group velocity. With the help of Eq.~\eqref{iota2}, we obtain for the diagonal component 
of the conductivity
\bea
\sigma_{jj}(\Ef)=e^2\tau_d (t_ja_j)^2\left[\nu(\Ef)+2\mathcal{I}(\Ef)\right],\label{cond2}
\eea
where $\nu(\Ef)$ is the density of states in Eq.~\eqref{DOSHC} evaluated at $E=\Ef$, and
\bea
\mathcal{I}(\Ef)= \Pi^{M-1}_{n=1} a_n^{-1} \int^\infty_0 d\rho \cos(\Ef\rho) J_2(t_j\rho) \Pi_{n=1,n\neq j}^{M-1} J_0(t_n\rho).\label{I}
\eea
From its definition in Eq.~\eqref{cond}, we see that $\sigma_{jj}(\Ef) \neq 0$ only if $|\Ef|<\sum_{n=1}^{M-1} t_n$. The first term in Eq.~\eqref{cond2} 
itself obeys this rule. 
Therefore, the integral \eqref{I} must also obey this rule.

In general, we consider the integral \eqref{g} with $\alpha=1$, subject to additional constraints: namely, $\nu_n$ are integers, such that i) $\sum_{n=1}^{M-1}\nu_n=2p$ and $B=0$ or ii) $\sum_{n=1}^{M-1}\nu_n=2p+1$ and $A=0$, with $p=0,1,2\dots$.\ In both cases, the product of the Bessel functions is an even function of $\rho$, and the integral over $\rho $ can be extended over the entire real axis. By working backwards through steps in the previous examples, we obtain for case i),
\bea
g(\{c_n\})&=&\frac{1}{2} \R \int^\infty_{-\infty} d\rho e^{-ic_M\rho}\Pi_{n=1}^{M-1}\int \frac{dk_n}{2\pi i^{\nu_n}} e^{i\left(\rho c_n\cos k_n +\nu_n k_n\right)}\nn \\
&&\hspace{-.3cm}=\frac{(-1)^p}{2} \Pi_{n=1}^{M-1}\int \frac{dk_n}{2\pi}
\delta\left(c_M-\sum_{n=1}^{M-1} c_n\cos k_n\right)\cos\left(\sum_{n=1}^{M-1} \nu_n k_n\right).\label{g2}
\eea
The polygonal constraint now follows simply from $c_M=\sum_{n=1}^{M-1} c_n\cos k_n\leq \sum_{n=1}^{M-1} c_n$. Case ii) is analyzed in the same way, with $\R$ in Eq.~\eqref{g2} replaced by $\I$.

\section{\label{ref:phys motiv}Umklapp scattering of Bloch electrons} 
\subsection{\label{ref:subsec two particle}Example: Two-particle umklapp scattering on a honeycomb lattice} 
\begin{figure}[!]
  \subcaptionbox{\label{fig:recip lattice}}{\includegraphics[scale=0.45]{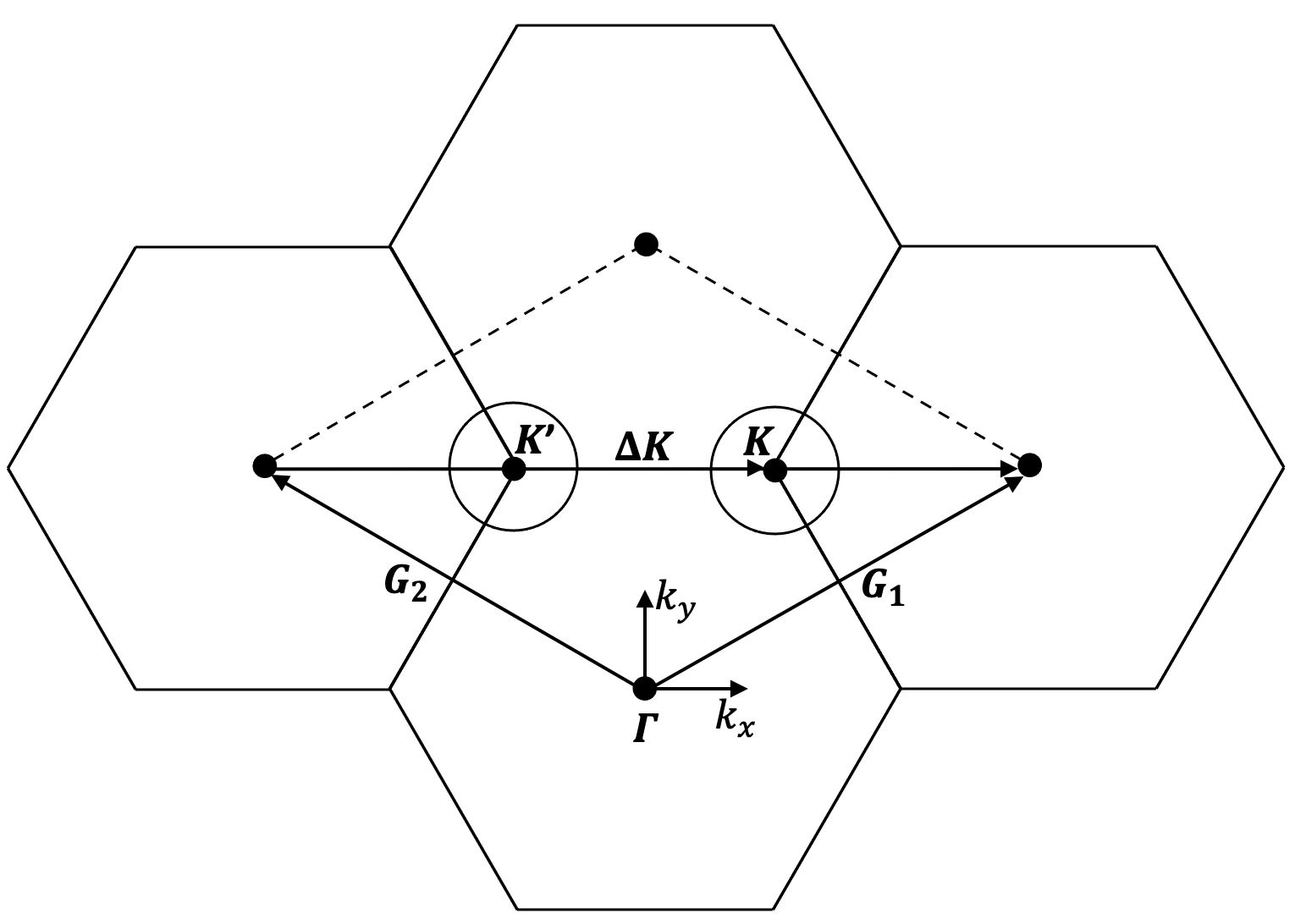}}
  \hfill
  \subcaptionbox{\label{fig:2part}}{\includegraphics[scale=0.7]{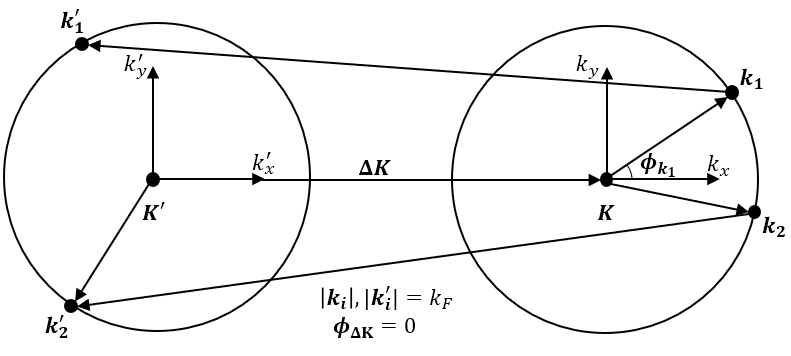}}
  \caption{(a) The reciprocal lattice of a real-space honeycomb lattice in real space is also a honeycomb lattice. 
 $\bK$ and $\bK'$ points are the degeneracy points of the electron spectrum. The Brillouin zone is defined by the region enclosed by $\bG_1$ and $\bG_2$ and the two dashed lines.  (b) An example of a two-particle intervalley scattering process: electrons with initial momenta $\bk_1$ and $\bk_2$ (measured relative to the $\bK$ point) scatter 
 into the states with momenta $\bk_1'$ and $\bk_2'$ (measured relative to the $\bK'$ point) on the 
 Fermi surface around the $\bK'$ point.  A scattering process can lead to an umklapp only if 
 $4\kf\ge|\Delta\bK|$.
 The model is applicable as long as the Fermi surfaces do not overlap, i.e., for 
$2\kf<|\Delta\bK|$.}  
\end{figure}

\begin{figure}[!]
  \subcaptionbox{\label{fig:hc dispersion}}{\includegraphics[scale=0.58]{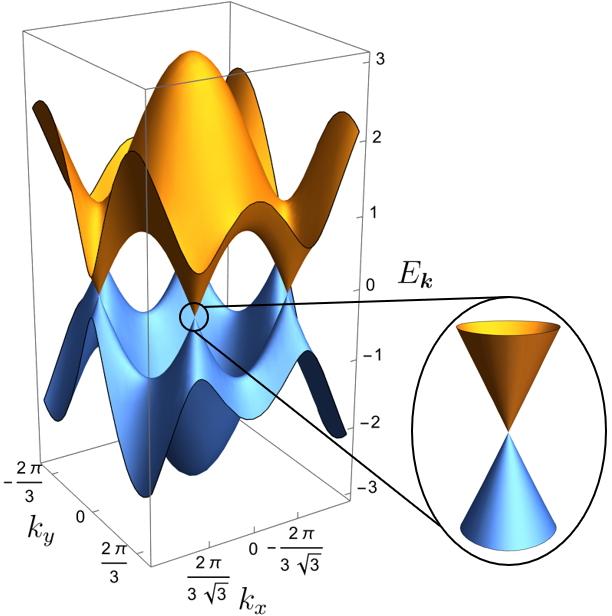}}
  \hfill
  \subcaptionbox{\label{fig:hc fs}}{\includegraphics[scale=.5]{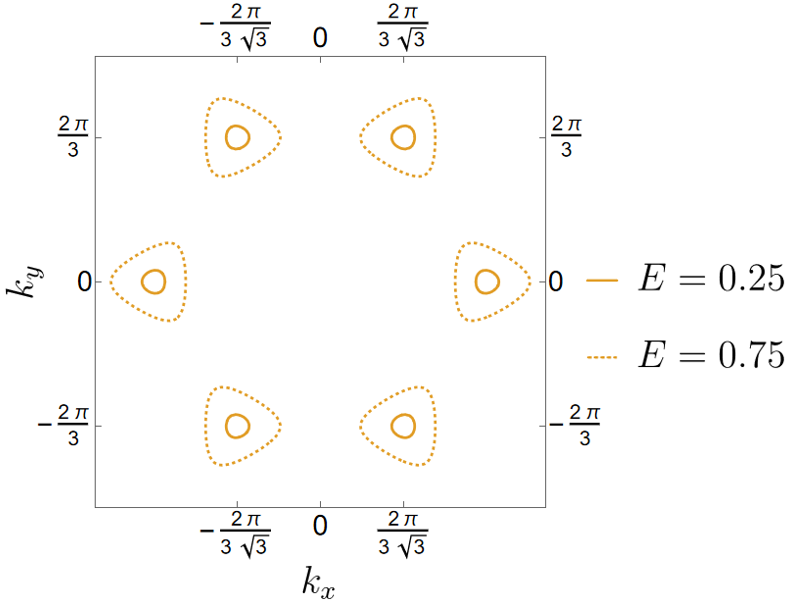}}
  \caption{(a) The full dispersion of electrons in graphene obtained from the nearest-neighbor tight-binding model \cite{CastroNeto2009TheGraphene}. Encircled is a zoomed in view of the Dirac cones (with finite trigonal warping) centered about a $\bK$ point. 
  (b) Fermi surface contours for two different energies (in units of the hopping parameter).}
\end{figure}

To provide the physical context for the integral \eqref{alpha}, we start with a simple example of two-electron scattering on a 2D honeycomb lattice, encountered, e.g., in graphene. As depicted in Fig.~\ref{fig:recip lattice}, the reciprocal lattice is also a honeycomb. The first Brillouin zone 
may be defined by the reciprocal lattice vectors $\bG_1$ and $\bG_2$ and contains two nonequivalent corner points, $\bK$ and $\bK'$. A special feature of the honeycomb lattice is that the electron energy spectrum is doubly degenerate (in addition to spin degeneracy) at the corner points. Near these points, the spectrum is that of massless Dirac fermions:  $\ve_\bk=\pm \vf k$, where $\vf$ is the ($k$-independent) Fermi velocity and $\pm$ corresponds to the conduction/valence bands \cite{CastroNeto2009TheGraphene}. The full dispersion of electrons in graphene
and a zoomed view of the Dirac cones are depicted in Fig.~\ref{fig:hc dispersion}. (The lattice constant and hopping parameter were set to unity). In Fig.~\ref{fig:hc fs}, the isoenergetic contours are plotted at two different energies.  At lower energies, the Fermi surfaces are nearly isotropic, while at higher energies the contours display more noticeable trigonal warping.

If the Fermi energy is either above or below the touching points of two Dirac cones, the system is a metal with a Fermi surface that consists of two  contours ("valleys") centered  at the $\bK$ and $\bK'$ points. 
(Without loss of generality, we will assume that the Fermi level is at energy $\Ef$ above the touching point.) At lower fillings, when the size of the valleys is much smaller than the reciprocal lattice vector, the valleys are almost circular, as depicted in Fig.~\ref{fig:recip lattice}.
At higher fillings, the valleys get (trigonally) distorted; however, we are going to ignore this in our simple model.
As we said in Sec.~\ref{sec:intro}, umklapp events require large changes in the total quasimomentum of the colliding electrons. For a two-valley system, the largest change in the momentum is achieved when  two electrons are transferred simultaneously from one valley to another. Simple geometry shows that the vector $\Delta\bK$ connecting the $\bK$ and $\bK'$ points is given by $\Delta\bK=\bK-\bK'=(\bG_1-\bG_2)/3$. On the other hand,  any superposition of the lattice vectors is also a lattice vector, and thus $\bG=\bG_1-\bG_2=3\Delta\bK$ is a lattice vector as well. 
If the electron momenta in each valley are measured from the valley centers, 
a two-electron intervalley transfer becomes an umklapp event if
\bea
\bk_1+\bk_2-\bk'_1-\bk'_2+2(\bK-\bK')>\bG_1-\bG_2
\label{umklapp condition}
\eea
or
\bea
\bk_1+\bk_2-\bk'_1-\bk'_2>\Delta\bK.
\label{umklapp condition}
\eea
As 
shown in Fig.\ \ref{fig:2part}, we consider two 
electrons initially in the $\bK$ valley that scatter to the $\bK'$ valley. For circular valleys of radii $\kf$, the magnitude of the left-hand
side of Eq.~\eqref{umklapp condition} is maximized when two electrons are transferred from, say, the rightmost edge of valley $\bK$ to the leftmost edge of valley $\bK'$, such that $\left\vert\bk_1+\bk_2-\bk'_1-\bk'_2\right\vert=4\kf$. Therefore, umklapps are possible for $\kf>\Delta K/4=G/12$, i.e., when $\kf$ is still sufficiently small compared to $G$. {\it A posteriori}, this serves as a justification (albeit only a numerical one) 
for neglecting the trigonal distortion of the Fermi surface. 
On the other hand, the valleys touch each other at $\kf=\Delta K/2$, after which the model becomes meaningless. (For a real honeycomb lattice, the Fermi surface undergoes a reconstruction at some critical filling, when a new pocket is opened at the center of the Brillouin zone).
Thus, the relevant range for umklapp scattering is $\Delta K/4\le \kf< \Delta K/2$.

The scattering rate of an electron with momentum $\bk_1$ due to an umklapp process is given by the Fermi Golden Rule 
 \bea
  \frac{1}{\tau_{\bk_1}}&=&\int\frac{d^2k_2}{(2\pi)^2}\int\frac{d^2k_1'}{(2\pi)^2}\int\frac{d^2k_2'}{(2\pi)^2}\Wt\left[n_{\bk_2}(1-n_{\bk_1'})(1-n_{\bk_2'})+n_{\bk_1'}n_{\bk_2'}(1-n_{\bk_2})\right]\nn\\
  &&\times \delta(\bk_1+\bk_2-\bk_1'-\bk_2'-\Delta\bK)\delta(\ve_{\bk_1}+\ve_{\bk_2}-\ve_{\bk_1'}-\ve_{\bk_2'}),
  \label{FGR}
 \eea
 where $\Wt$ is the scattering 
 probability density, $n_\bk$ is the Fermi function, and $\ve_\bk$ is the electron energy. 
For our purposes, the screened Coulomb interaction between electrons can be replaced by a $\delta$-function potential with amplitude $
u$. In this case, we have (see Secs.~I-II of the supplementary material for details)
\begin{equation}
\Wt=16\pi 
u^2\sin^2\left(\frac{\phi_1+\phi'_1}{2}\right)\sin^2\left(\frac{\phi_2+\phi'_2}{2}\right),\label{W}
\end{equation}
where $\phi_1$ is the azimuthal angle of $\bk_1$, etc. 
The angular dependence of $\Wt$ comes from the spinor structure of the wavefunctions of the Dirac Hamiltonian. It is convenient to choose $|\bk_1|=\kf$ and average $1/\tau$ over the directions of $\bk_1$. In doing so, we obtain for the averaged scattering rate at $T\ll \Ef$ ,
\bea
\left\langle \frac{1}{\tau}\right\rangle =\frac{\pi}{2}
\lambda^2
\frac{T^2}{\Ef}f\left(c\right),\label{avtau}
\eea
where $\lambda=\kf u/\vf$ is a dimensionless coupling constant,  $c=\Delta K/\kf$, $f(c)=S_1(c)-2S_2(c)+S_3(c)$, and
\begin{flalign}
\begin{split}
S_j(c)
= \frac{1}{(2\pi)^{5}}\int_0^{2\pi} d\phi'_1& \int_0^{2\pi} d\phi'_2 \int_0^{2\pi} d\phi_1 \int_0^{2\pi} d\phi_2 
\ \delta(\cos{\phi'_1}+\cos{\phi'_2}-\cos{\phi_1}-\cos{\phi_2}-c
\cos\phi_{\Delta\bK}) \\
&\hspace{2.8cm}\times \ \delta(\sin{\phi'_1}+\sin{\phi'_2}-\sin{\phi_1}-\sin{\phi_2}-c\sin\phi_{\Delta\bK})R_j,\ \;j=1,2,3 \label{zeta1}\\
\end{split}
\end{flalign}
with $R_1=1$, $R_2=\cos(\phi_1+\phi_1')$, and $R_3=\cos(\phi_1+\phi_1')\cos(\phi_2+\phi_2')$.
In the last equation, the $\delta$-functions of momentum conservation were re-written in terms of the Cartesian components of the total quasimomentum. Although in our geometry vector $\Delta\bK$ has only an $x$-component, we allowed also for possible $y$-components for the sake of generality. This completes the prelude from condensed matter physics.

\subsection{\label{sec:N=2}Integrals of products of Bessel functions for two-particle umklapp scattering}
Now we will show that the integral in Eq.~\eqref{zeta1} is, indeed, a special case of Eq.~\eqref{alpha}. First, we re-write the $\delta$-functions in Eq.~\eqref{zeta1} as
\beq
\delta(r)=\int_{-\infty}^{\infty} dx e^{ixr},
\eeq
such that 
\begin{flalign}
\begin{split}
S_j(c)&= \frac{1}{(2\pi)^{5}}\int_{-\infty}^{\infty} dx \int_{-\infty}^{\infty} dy \int_0^{2\pi} d\phi'_1 \int_0^{2\pi} d\phi'_2 \int_0^{2\pi} d\phi_1 \int_0^{2\pi} d\phi_2 \\  &\hspace{.04\textwidth} \times \ e^{ix(\cos{\phi'_1}+\cos{\phi'_2}-\cos{\phi_1}-\cos{\phi_2}-c\cos\phi_{\Delta\bK})} \ e^{iy(\sin{\phi'_1}+\sin{\phi'_2}-\sin{\phi_1}-\sin{\phi_2}-c\sin\phi_{\Delta\bK})}R_j. \label{theta}
\end{split}
\end{flalign}
The Cartesian 
coordinates may be transformed to polar 
ones
such that $x=\rho\cos\varphi$ and \\ $y=\rho\sin\varphi$, giving
\beq
x\cos{\phi}+y\sin\phi=\rho(\cos{\varphi}\cos{\phi}+\sin{\varphi}\sin{\phi}) = \rho\cos(\phi-\varphi)
\eeq
for all $\phi=\phi_1,\dots ,\phi_2',\phi_{\Delta\bK}$. For $S_1$ with $R_1=1$, we then obtain
\begin{flalign}
\begin{split}
S_1(c)&= \frac{1}{(2\pi)^{5}}\int_{0}^{\infty} d\rho\rho \int^{2\pi}_0 d\varphi \int_0^{2\pi} d\phi'_1 \int_0^{2\pi} d\phi'_2 \int_0^{2\pi} d\phi_1 \int_0^{2\pi} d\phi_2 \\  &\hspace{.04\textwidth} \times \ 
e^{i\rho\left(\cos(\phi_1'-\varphi)+\cos(\phi_2'-\varphi)-\cos(\phi_1-\varphi)-\cos(\phi_2-\varphi)-c\cos(\phi_{\Delta\bK}-\varphi)\right)}. \label{theta}
\end{split}
\end{flalign}
The integrations over $\phi_1\dots \phi_2'$ are carried out independently from each other as each such integration goes over the full period of the integrand and therefore does not depend on $\varphi$. Using  the integral representation of a Bessel function of integer order, Eq.~\eqref{iota2}, we have
\bea
S_1(c)=\frac{1}{2\pi}\int_{0}^{\infty} d\rho\rho J_0^4(\rho)\int^{2\pi}_0 d\varphi e^{-ic\rho\cos(\phi_{\Delta\bK}-\varphi)}.
\eea
Applying Eq.~\eqref{iota2} again, we obtain
\beq
S_1(c) = \int_{0}^\infty d\rho\rho J_0^4(\rho)J_0(c\rho).\label{s1}
\eeq
 The remaining two integrals in Eq.~\eqref{zeta1}, $S_2$ and $S_3$, differ from $S_1$ only in the additional angular dependence of factors $R_2$ and $R_3$. Following the same steps as before, we obtain
\begin{flalign}
\begin{split}
S_2(c)&= -\cos(2\phi_{\Delta \bK})\int_0^{\infty} d\rho\rho J_0^2(\rho) J_1^2(\rho) J_2(c\rho)
\label{s2}
\end{split}
\end{flalign}
and
\begin{flalign}
\begin{split}
S_3(c)
&= \frac{\cos(4\phi_{\Delta\bK})}{2}\int_0^{\infty} d\rho\rho J_1^4(\rho)J_4(c\rho)+\frac{1}{2}\int^\infty_0 d\rho\rho J_1^4(\rho) J_0(c\rho).
\label{s3}
\end{split}
\end{flalign}

We see that Eqs.~\eqref{s1}-\eqref{s3} are indeed special cases of Eq.~\eqref{alpha} with $M=5$, $\alpha=2$, $c_1=\dots=c_4=1$, and $c_5=c$, such that the corresponding polygon, defined by Eq.~\eqref{PC}, is a pentagon with the four shorter sides being of the same length.  The indices of the Bessel functions in all three equations are such that the charge-neutrality condition \eqref{CNC} is satisfied.
For $S_1$, $\nu_1=\dots=\nu_5=0$ such that Eq.~\eqref{CNC} is satisfied with $m=1$:
\beq 
\nu_5=0= 4\times0 + 2 - 2\times 1.
\eeq
For $S_2$, $\nu_1=\nu_2=0$, $\nu_3=\nu_4=1$, and $\nu_5=2$,  such that Eq.~\eqref{CNC} is also satisfied with $m=1$
\beq 
\nu_5=2=2\times(0+1)+2-2\times 1.
\eeq
Likewise, it is easy to check that for the first and second terms of $S_3$, Eq.~\eqref{CNC} is satisfied with $m=1$ and $m=3$, respectively.

Since Eq.~\eqref{CNC} is satisfied, then, according to Eq.~\eqref{alpha1}, all the integrals in Eq.~\eqref{s1}-\eqref{s3} must vanish if Eq.~\eqref{PC} is \emph{not} satisfied, i.e., if $c>4$. Now we see why this is the case. Indeed, in our physical model $c=\Delta K/\kf$, hence $c>4$ coincides with the condition $\kf<\Delta K/4$, when uklapp scattering is \emph{forbidden}, and thus $f(c)$ in Eq.~\eqref{avtau} must be equal to zero. This physical condition is enforced by the first $\delta$-function in Eq.~\eqref{FGR}, 
and thus 
the
somewhat mysterious mathematical property encoded by Eq.~\eqref{alpha1} allows for a simple interpretation in terms of momentum conservation.  
\begin{figure}[!]
    \includegraphics[scale=1]{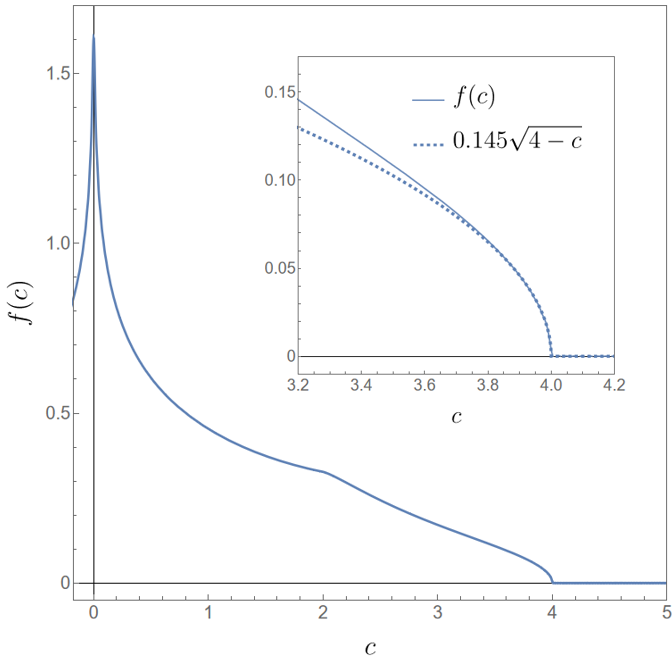}
    \caption{Main panel: Function $f(c)$ entering the average scattering rate due to two-particle umklapp scattering on a honeycomb lattice, Eq.~\eqref{avtau}. In this example, the polygonal constraint is satisfied if $c<4$. Additional discontinuities in the derivative of $f(c)$ at $c=0$ and $c=2$ 
    are discussed in Sec.~III of the supplementary material. (Although the physical values of $c$ are positive, we also showed a small region of negative $c$ in order to emphasize the discontinuity of the derivative at $c=0$.)
    Inset: closer view of  
    $f(c)$ near the threshold at $c=4$. As discussed in Sec.\ \ref{ref:subsec thresh behav}, $f(c)$ scales as $\sqrt{4-c}$ in the limit $c\rightarrow 4$ from below. 
    The dashed line is a fit by the expression displayed in the legend.
    }
    \label{fig:N eq 2 plot}
\end{figure}

The function $f(c)$ is plotted in Fig.~\ref{fig:N eq 2 plot} for $\phi_{\Delta\bK}=0.$ (Although our physical model makes sense only for $c>2$, when the valleys are separate, there are no mathematical constraints on the value of $c$.) In addition to the expected discontinuity of $df/dc$ at $c=4$, there are two other discontinuities at $c=0$ and $2$, the latter of which corresponds to the Fermi surfaces first touching. The discontinuities in the derivative are analyzed in Sec.~III of the supplementary material. We will discuss the threshold behavior of $f(c)$ in Sec.~\ref{ref:subsec thresh behav},  after generalizing our results for the case of $N$-particle scattering in the next section.

\subsection{\label{ref:subsec gen higher order}$N$-particle scattering on an abstract 2D lattice}
\begin{figure}[t]
    \centering
    \includegraphics[scale=0.5]{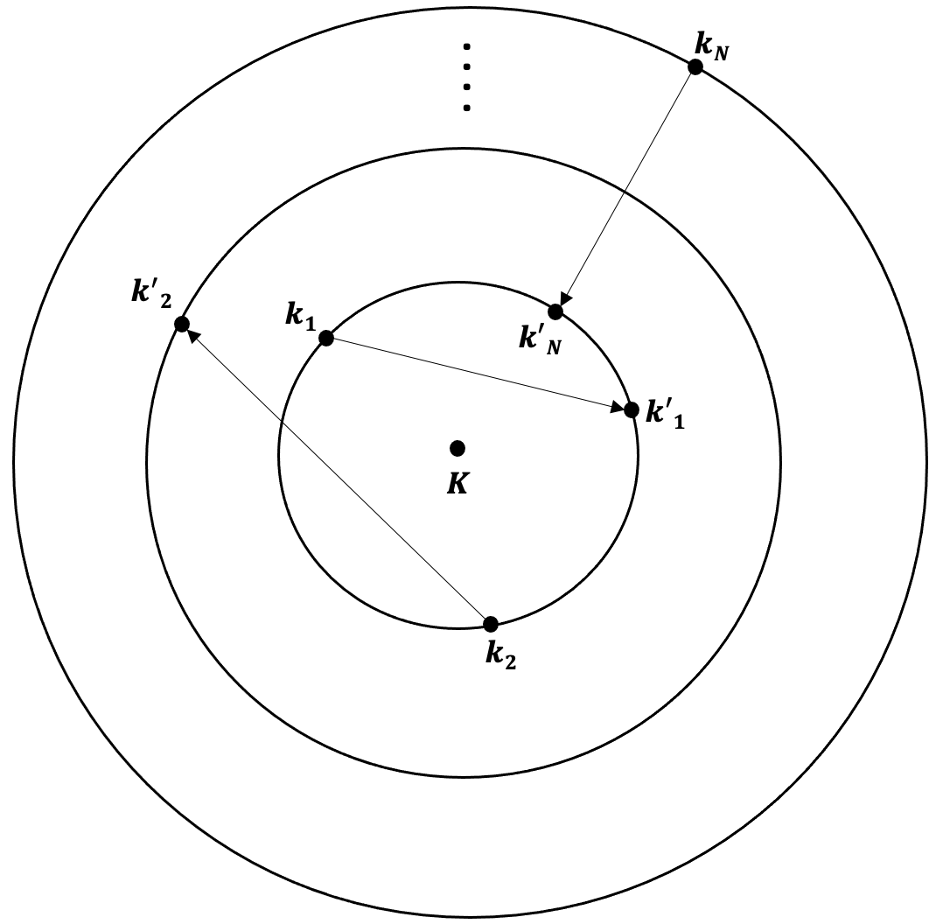}
    \caption{ 
    An example of an N-particle scattering event
    on a 2D 
    Fermi surface, consisting of concentric circles.
    The $N$ initial 
    momenta 
    $\bk_{1...N}$ and $N$ final momenta
    $\bk'_{1...N}$ reside on concentric Fermi surfaces, with some possibly sharing the same Fermi surface.
    }
    \label{fig:concentric}
\end{figure}

In the previous section, we considered a particular model of Dirac fermions with two nonequivalent degeneracy points, which is the low-energy model of graphene. Some features of this model, such as the additional angular dependence of the scattering probability in Eq.~\eqref{W}, are specific for a given Hamiltonian. In this section, we take a more abstract approach and focus only on the constraints imposed by momentum conservation for an $N$-particle umklapp scattering process, without specifying a particular lattice model.
We consider a hypothetical 2D Fermi surface consisting of $N_{\mathrm{FS}}$ concentric circles with radii $k_{\mathrm{F},1}\dots k_{\mathrm{F},N_{\mathrm{FS}}}$, as shown in Fig.~\ref{fig:concentric}, 
and examine the case when $N$ electrons, with initial momenta on any of the circles, scatter into their final states, also with momenta on any of the circles, such that the change in momentum matches the shortest reciprocal lattice vector $\bG$. The initial and final states may be on the same or different circles. 
Momentum conservation is enforced via a $\delta$-function
\bea
\delta\left(\sum_{n=1}^N\left(\bk'_n-\bk_n\right)-\bG\right),\label{Np}
\eea
where the magnitude of each momentum is equal to the radius of one of the circles.
The momenta may be rescaled to dimensionless ones such that $c_n=k'_{
n}/k_{\mathrm{F},1}$ for odd $n$, $c_n=k_{
n}/k_{\mathrm{F},1}$ for even $n$, and $c_{2N+1}=G/k_{F1}$. Since the magnitude of each final and initial momentum is equal to one of the Fermi surface radii, $c_n=k_{\mathrm{F},n}/k_{\mathrm{F},1}$ both for even and odd $n$.
If the scattering probability $W_{\bk_1,\dots,\bk_N;\bk'_1,\dots,\bk'_N}$ is replaced by a constant,
the average scattering rate is proportional to the function
\begin{equation}
\begin{split}
f(\{c_n\}
)=\frac{1}{(2\pi)^{2N+1}}\prod_{n=1}^{2N}\int^{2\pi}_0 d\phi_{n} \  \delta\left(\sum_{n=1}^{2N}(-1)^{n-1}c_{n}\cos\phi_{n}-c_{2N+1}\cos\phi_{2N+1}\right) \\
\times \ \delta\left(\sum_{n=1}^{2N}(-1)^{n-1}c_{n}\sin\phi_{n}-c_{2N+1}\sin\phi_{2N+1}\right)\label{kappa}
\end{split}
\end{equation} 
with $\{c_n\}=c_1,\dots, c_{2N+1}$.
We can then follow the same procedure as for the two-particle case, i.e.,
replace the $\delta$-functions by their integral forms and interchange the order of integration
\bea
f(\{c_n\}
)&=&\frac{1}{(2\pi)^{2N+1}} \int_{-\infty}^{\infty} dx \int_{-\infty}^{\infty} dy \prod_{n=1}^{2N}\int^{2\pi}_0 d\phi_{n} \ e^{ix\left(\sum_{n=1}^{2N}(-1)^{n-1}c_{n}\cos\phi_{n}-c_{2N+1}\cos\phi_{2N+1}\right)}\nn \\ && \hspace{5.3cm} \times \ e^{iy\left(\sum_{n=1}^{2N}(-1)^{n-1}c_{n}\sin\phi_{n}-c_{2N+1}\sin\phi_{2N+1}\right)}. \label{lambda}
\eea
Like arguments in the exponential terms may once again be combined and independently integrated as
\beq
\frac{1}{2\pi}\int_0^{2\pi} d\phi_{n} e^{\pm ic_n(x\cos{\phi_n}+y\sin{\phi_n})} = J_0(c_n\rho),
\eeq
leading to a product of Bessel functions, so that Eq.~\eqref{lambda} is simplified to 
\beq
f(\{c_n
\}
)=\frac{1}{2\pi} \int_{0}^{\infty} d\rho \rho  \prod_{n=1}^{2N} J_0(c_n\rho) \int^{2\pi}_0 d\varphi \ e^{-ic_{2N+1}(x\cos\phi_{2N+1}+y\sin\phi_{2N+1})} 
\eeq
The integration over $\varphi$ reduces the last expression to the final form 
\beq
f(\{c_n\})=\int^\infty_0 d\rho \rho \prod_{n=1}^{2N+1}J_{0}(c_n\rho), \label{lambda2}
\eeq
which is again a special case of Eq.~\eqref{alpha} with $M=2N+1$, $\alpha=2$, and $\nu_{1}=\dots=\nu_{2N+1}=0$. The charge neutrality condition is clearly satisfied with $m=1$, so that when the polygonal constraint is broken, namely $c_{2N+1}>\sum_{n=1}^{2N}c_n$, the integral vanishes. In terms of physical variables, the latter condition reads $\sum_{n=1}^{2N} k_{\mathrm{F},n}<G$, which means that the Fermi surfaces are too small to support an umklapp process. Again, the polygonal constraint is ensured by the structure of the $\delta$-functions in Eqs.~\eqref{Np} and \eqref{kappa}.

Bessel functions of non-zero orders arise due to the additional angular dependence of the scattering probability, $W_{\bk_1,\dots,\bk_N;\bk'_1,\dots,\bk'_N}$. In the realm of physical reality, such a dependence comes in the form of integer powers of trigonometric functions, $\cos\phi_n$ and $\sin\phi_n$, or, more generally, can be represented by infinite series of such functions. As in the two-particle case,  additional factors of $\cos\phi_n$ and $\sin\phi_n$ will promote at least some of the indices of the Bessel functions to non-zero values. However, the indices will remain integer (or half-integer in 3D, cf. Sec.~\ref{ref:scatt 3d}). 

\subsection{\label{ref:subsec thresh behav} Threshold Behavior} 
If the polygonal constraint is satisfied, a closed-form result for the integral \eqref{alpha} is not known, except for some special cases. However, a representation of this integral in terms of $\delta$-functions allows one to find  
 the asymptotic behavior of the function $f(\{c_n\})$ just below  the threshold
for which the polygonal constraint \eqref{PC} is no longer satisfied,
i.e., for
 $c_{2N+1}=\sum_{n=1}^{2N}c_n-\epsilon$ with $0<\epsilon\ll 1$. Setting $\phi_{2N+1}=0$ for simplicity, we obtain instead of Eq.~\eqref{kappa}
\begin{flalign}
\begin{split}
f(\{c_n\})&=\frac{1}{(2\pi)^{2N+1}}\prod_{n=1}^{2N}\int^{2\pi}_0 d\phi_{n}  \delta\left(\epsilon -\sum^{2N}_{n=1}c_n\left[1-(-)^{n-1}\cos\phi_n\right]
\right)
\delta\left(\sum_{n=1}^{2N}c_{n}(-)^{n-1}\sin\phi_{n}\right).\label{nearth1}
\end{split}
\end{flalign}
For $\epsilon\ll 1$,  the support of the integrals over $\phi_n$ 
comes from narrow regions near  $0$ (for odd $n$) and near $\pi$ (for even $n$).
Expanding $\cos\phi_{2p+1}=1-\phi^2_{2p+1}/2$, $\cos\phi_{2p}=\cos(\pi-\bar\phi_{2p})=-1+\bar\phi^2_{2p}/2$, $\sin\phi_{2p+1}=\phi_{2p+1}$, $\sin\phi_{2p}=\bar\phi_{2p}$, and  extending the limits of integration to $\phi_{2p+1}\in(0,\infty)$ and $\phi_{2p}\in(-\infty,\infty)$,
we have 
\begin{flalign}
\begin{split}
 f(\{c_n\})
&\approx \frac{1}{
(2\pi)^{2N+1}}\prod_{p=0}^{
N-1}\int_0^\infty d\phi_{2p+1} \prod_{p=1}^{N}\int_{-\infty}^\infty d\bar\phi_{2p} \  \delta\left(\epsilon-\frac{1}{2}
\sum_{p=0}^{N-1}c_{2p+1}\phi_{2p+1}^2-\frac{1}{2}\sum_{p=1}^{N}c_{2p}\bar\phi_{2p}^2\right) \\ &\hspace{6.8cm} \times \delta\left(\sum_{p=0}^{N-1}c_{2p+1}\phi_{2p+1}-\sum_{p=1}^{N}c_{2p}\bar\phi_{2p}\right). \label{threshold form}
\end{split}
\end{flalign}
Rescaling the integration variables as $\phi_{2p+1}=\sqrt{\varepsilon}x$ 
and $\bar\phi_{2p}=\sqrt{\varepsilon}y$,
we see that $f(\{c_n\})$ 
behaves as
\beq
f(\{c_n\})\propto \ve^{N-\frac{3}{2}}=\Bigl(\sum_{n=1}^{2N}c_n-c_{2N+1}\Bigr)^{N-\frac{3}{2}}.
\label{nearth2}
\eeq
Note that a possible angular dependence of the scattering probability does not affect the threshold behavior in Eq.~\eqref{nearth2} 
because
the scattering probability can be evaluated at $\phi_{2p+1}=0$ and $\phi_{2p}=\pi$, which changes only the overall prefactor. 
For the case of two-particle scattering on a honeycomb lattice ($N=2$, $c_{1\dots 4}=1$, and $c_5=c$) considered in Secs.~\ref{ref:subsec two particle} and \ref{sec:N=2}, we have $f(\{c_n\})\propto \sqrt{\epsilon}=\sqrt{4-c}$. 
The inset in Fig.~\ref{fig:N eq 2 plot} shows that this is indeed the correct behavior near the threshold.

In the context of electron-electron scattering, $N\geq 2$. The $N=1$ case corresponds to the situation when an electron scatters off some other object, e.g., a phonon.
In this case, $f(\epsilon)\propto 1/\sqrt{\epsilon}$, i.e., the scattering rate diverges at the threshold.

\subsection{\label{ref:scatt 3d}Umklapp scattering on a 3D lattice} 
We now turn to $N$-particle scattering in 3D and show it provides a phyiscal interpretation of the vanishing of the integral \eqref{alpha} with Bessel functions of \emph{half-integer} orders, if Eq.~\eqref{CNC} is satisfied  but Eq.~\eqref{PC} is not.
The Fermi surfaces will now be concentric spheres.
Similar to the integral over circular Fermi surfaces found in \eqref{kappa}, the scattering rate will now be proportional to the following integral over spherical Fermi surfaces (for $\Wt=\mathrm{const}$):
\begin{flalign}
\begin{split}
f(\{c_n\})
&=\frac{1}{(4\pi)^{2N+1}}\prod_{n=1}^{2N}\int^\pi_0 d\theta_{n} \sin\theta_{n} \int^{2\pi}_0 d\phi_{n} \ \delta\left(\sum_{n=1}^{2N}(-1)^{n-1}c_{n}\sin\theta_{n}\cos\phi_{n}-c_{2N+1}\sin\theta_{2N+1}\cos\phi_{2N+1}\right) \\ 
&\hspace{6.4cm} \times \delta\left(\sum_{n=1}^{2N}(-1)^{n-1}c_{n}\sin\theta_{n}\sin\phi_{n}-c_{2N+1}\sin\theta_{2N+1}\sin\phi_{2N+1}\right) \\
&\hspace{6.4cm} \times \delta\left(\sum_{n=1}^{2N}(-1)^{n-1}c_{n}\cos\theta_n-c_{2N+1}\cos\theta_{2N+1}\right), \label{3dscat}
\end{split}
\end{flalign}
where $\theta_n$ and $\phi_n$ are the polar and azimuthal angles of vector $\bk_n$ with magnitude $k_{\mathrm{F},n}$. 
Upon converting the $\delta$ functions to integrals over complex exponentials and swapping the order of integration, the integral becomes
\begin{flalign}
\begin{split}
f(\{c_n\})=&\frac{1}{(4\pi)^{2N+1}}\int_{-\infty}^{\infty} dx \int_{-\infty}^{\infty} dy \int_{-\infty}^{\infty} dz \prod_{n=1}^{2N}\int^\pi_0 d\theta_n \sin\theta_n \int^{2\pi}_0 d\phi_n \  \\ 
&\hspace{8cm}\times e^{ix\left(\sum_{n=1}^{2N}(-1)^{n-1}c_n\sin\theta_n\cos\phi_n-c_{2N+1}\sin\theta_{2N+1}\cos\phi_{2N+1}\right)} \\ 
&\hspace{8cm}\times  e^{iy\left(\sum_{n=1}^{2N}(-1)^{n-1}c_n\sin\theta_n\sin\phi_n-c_{2N+1}\sin\theta_{2N+1}\sin\phi_{2N+1}\right)} \\
&\hspace{8cm}\times e^{iz\left(\sum_{n=1}^{2N}(-1)^{n-1}c_n\cos\theta_n-c_{2N+1}\cos\theta_{2N+1}\right)}. \label{rho}
\end{split}
\end{flalign}
Exponentials with like arguments may once again be combined and subsequently integrated over, where now integrals over both azimuthal and polar angles are present. By converting the Cartesian coordinates to spherical coordinates  $x=r\sin\vartheta\cos\varphi$, $y=r\sin\vartheta\sin\varphi$, and $z=r\cos\vartheta$, each of these $2N$ integrals may be written as
\begin{equation}
\begin{split}
I_n=\frac{1}{4\pi}\int_0^\pi d\theta_n \sin\theta_n \int_0^{2\pi} d\phi_n \ e^{\pm ic_n(x\sin\theta_n\cos{\phi_n}+y\sin\theta_n\sin{\phi_n}+z\cos\theta_n)} \\ 
= \frac{1}{4\pi} \int_0^\pi d\theta_n \sin\theta_n \int_0^{2\pi} d\phi_n \ e^{\pm ic_nr(\sin\vartheta\sin\theta_n\cos(\phi_n-\varphi)+\cos\vartheta\cos\theta_n)}. \label{sigma}
\end{split}
\end{equation}
The exponential terms may be separated and, due to $2\pi$ periodicity of the integrand in Eq.~\eqref{sigma}, the integral is reduced to
\begin{equation}
\begin{split}
I_n=\frac{1}{2} \int_0^\pi d\theta_n \sin\theta_n \ e^{\pm ic_nr\cos\vartheta\cos\theta_n} J_0(c_nr\sin\vartheta\sin\theta_n). 
\end{split}
\label{tau-1}
\end{equation}
Recalling the relation between the regular and spherical Bessel functions \cite{Sonine1880RecherchesSeries}, 
\beq
j_\nu(z)=\frac{1}{2\sin^\nu\vartheta}\int_0^\pi d\theta\sin^{\nu+1}\theta \ e^{ iz\cos\vartheta\cos\theta}J_\nu(z\sin\vartheta\sin\theta), \label{tau}
\eeq
we see that $I_n=j_0(c_n r)$, which leads us to the penultimate form
\begin{equation}
\begin{split}
f(\{c_n\})&=\int^\infty_0 dr r^2 \prod_{n=1}^{2N}j_{0}(c_n r) \ \frac{1}{2}\int_0^\pi d\vartheta\sin\vartheta \ e^{-ic_{2N+1}r\cos\vartheta\cos\theta_{2N+1}} \frac{1}{2\pi}\int_0^{2\pi} d\varphi \ e^{-ic_{2N+1}r\sin\vartheta\sin\theta_{2N+1}\cos\varphi}.
\end{split}
\label{jJ}
\end{equation}
Applying Eq.~\eqref{tau} again, we arrive at the final form
\beq
f(\{c_n\})=\int^\infty_0 dr r^2 \prod_{n=1}^{2N+1}j_{0}(c_n r).\label{tau3}
\eeq
Recalling the additional definition relating spherical Bessel functions of order $\nu$ with regular Bessel functions of order $\nu+1/2$ 
\bea
j_{\nu}(z)=\sqrt{\frac{\pi}{2z}}J_{\nu+1/2}(z),
\eea
it is clear that \eqref{tau3} can be rewritten in the same form as \eqref{alpha} with $\alpha=3-(2N+1)/2$.
If the polygonal constraint is not satisfied, 
the integral \eqref{3dscat} vanishes independently from the values of $\theta_{2N+1}$ and $\phi_{2N+1}$ so that they may be set to $\pi/2$ and zero respectively.
It is then clear that the same polygonal constraint is given by the inequality $c_{2N+1}\le\sum_{n=1}^{2N}c_n$, which is the same as for the 2D case. This explains why the integral in Eq.~\eqref{tau3} vanishes identically when the polygonal constraint is not satisfied.
As in the 2D case, spherical Bessel functions of non-zero order occur due to the angular dependence of the scattering probability; however, we will not dwell on this issue here.

\section{\label{ref: math conseq}Higher Dimensionalities} 
In the previous sections, we considered a special case of the integral \eqref{alpha} for $\alpha=2$ which, in a physical interpretation, corresponds to the 2D case. In this section, we show that the same analysis can be extended for any even, integer dimensionality, i.e.,  for $\alpha=2p$, $p=1,2\dots$

Consider the following integral
\beq
f(\{c_n\})=\int^\infty_0 d\rho \rho^{2p-1} \prod_{n=1}^{M>1}J_{0}(c_n \rho). \label{upsilon}
\eeq
Applying the same transformations as before, we obtain
\bea
f(\{c_n\})&=&
\frac{1}{(2\pi)^{M}} \prod_{n=1}^{M-1} \int_0^{2\pi} d\phi_n \int_{-\infty}^{\infty} dx \int_{-\infty}^{\infty} dy \ \bigl(x^2+y^2\bigr)^{
p-1} \ e^{ix(\sum_{n=1}^{M-1}c_n\cos\phi_n-c_M\cos\phi_M)}  e^{iy(\sum_{n=1}^{M-1}c_n\sin\phi_n-c_M\sin\phi_M)}\nn \\
&&\hspace{-.31cm}=\frac{1}{(2\pi)^{M}} \prod_{n=1}^{M-1} \int_0^{2\pi} d\phi_n \int_{-\infty}^{\infty} dx \int_{-\infty}^{\infty} dy \sum_{l=0}^{p-1} C_{p-1}^l x^{2l} y^{2(p-1-l)}
e^{ix(\sum_{n=1}^{M-1}c_n\cos\phi_n-c_M\cos\phi_M)}\nn\\
&& \hspace{9.2cm}\times \ e^{iy(\sum_{n=1}^{M-1}c_n\sin\phi_n-c_M\sin\phi_M)},
\label{phi}
\eea
where we used the binomial expansion  at the last step with $C^m_n=n!/m!(n-m)!$ being the binomial coefficient.
Relabeling $x_M=c_M\cos\phi_M$ and $y_M=c_M\sin\phi_M$,
we rewrite $x^{2l}$ as the $(2l)^\mathrm{th}$ derivative of the first exponential factor $x^{2l}...=(-)^l\partial^{2l}_{x_M}\dots$, and similarly for $y^{2(p-1-l)}$. Integrating over $x$ and $y$, we obtain 
\begin{flalign}
\begin{split}
f(\{c_n\})&= \frac{1}{(2\pi)^M} \prod_{n=1}^{M-1} \int_0^{2\pi} d\phi_n \sum_{l=0}^{p-1} C_{p-1}^l(-)^{p-1} \partial^{2l}_{x_M}\partial^{2(p-1-l)}_{y_M}
\delta\left(\sum_{n=1}^{M-1}c_n\cos\phi_n-x_M\right)\delta\left(\sum_{n=1}^{M-1}c_n\sin\phi_n-y_M\right).
\end{split}
\end{flalign}
If the polygonal constraint is not satisfied, each of the $\delta$-functions in the equation above vanishes identically along with all its derivatives.

\begin{figure}[t]
    \centering
    \includegraphics[scale=.65]{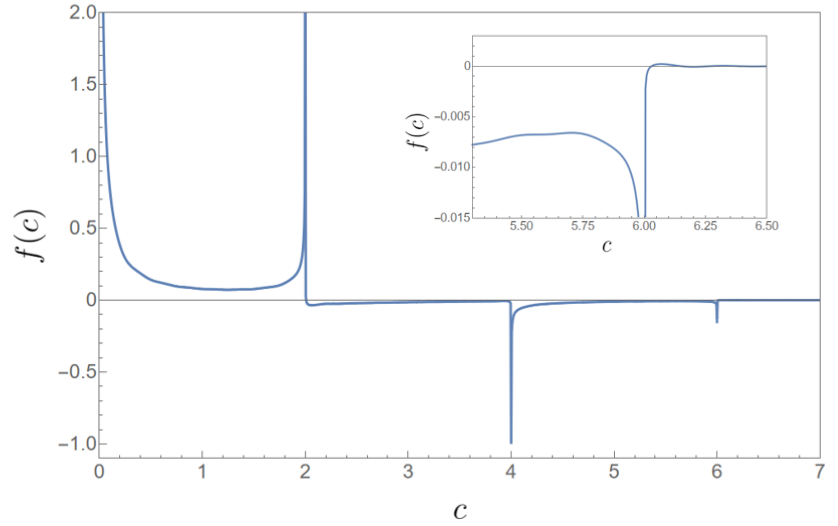}
    \caption{Numerical result for the integral \eqref{Meq7}.
    }
    \label{fig:Meq7}
\end{figure}

As an example, consider the integral \eqref{upsilon} with $p=2$ and $M=7$ (the smallest $M$ for which the integral with $\alpha=4$ is convergent)
\beq
f(c)=\int_0^{\infty} d\rho \rho^{3} J_0^6(\rho)  J_0(c\rho).
\label{Meq7}
\eeq
With all $\nu_n=0$ and $\alpha=4$, the charge neutrality condition \eqref{CNC} is satisfied with $m=2$, and thus the integral vanishes when the polygonal constraint \eqref{PC} is not satisfied, namely for $c>6$. The numerical result for Eq.~\eqref{Meq7} is shown in Fig.~\ref{fig:Meq7}, where it is seen 
$f(c)$ is indeed nonzero only when $c<6$, within a reasonable numerical accuracy.

The analysis above can be generalized for Bessel functions of arbitrary integer orders, in which case
\begin{flalign}
\begin{split}
f(\{c_n\})&= \int_0^\infty d\rho \rho^{2p-1} \prod_{n=1}^{M-1} \int_0^{2\pi} \frac{d\phi_n}{2\pi i^{\nu_n}} \ e^{i\sum_{n=1}^{M-1}(c_n\rho\cos\phi_n+\nu_n\phi_n)}  \frac{(-1)^{\nu_M}}{2\pi i^{\nu_M}}\int_0^{2\pi} d\varphi \ e^{i(-c_M\rho\cos\varphi+\nu_M\varphi)}. 
\end{split}
\end{flalign}
Upon transforming to Cartesian coordinates 
and shifting the variables as
$\varphi \rightarrow \varphi-\phi_M$ and 
$\phi_n \rightarrow \phi_n-\varphi$, 
the integral is reduced to
\begin{flalign}
\begin{split}
f(\{c_n\})&=e^{-i\nu_M\phi_M} \frac{(-1)^{\nu_M}}{2\pi i^{\nu_M}}\prod_{n=1}^{M-1}\int_0^{2\pi}  \frac{d\phi_n}{2\pi i^{\nu_n}} \ e^{ i\sum_{n=1}^{M-1}\nu_n\phi_n} \\
& \times \ \int_{-\infty}^{\infty} dx \int_{-\infty}^{\infty} dy \ \bigl(x^2+y^2\bigr)^{p-1}  \ e^{ix(\sum_{n=1}^{M-1}c_n \cos\phi_n-c_M\cos\phi_M)} e^{iy(\sum_{n=1}^{M-1}c_n \sin\phi_n-c_M\sin\phi_M)} e^{i (\nu_M - \sum_{n=1}^{M-1}\nu_n)\varphi},
\end{split}
\end{flalign}
If the charge neutrality condition \eqref{CNC} is satisfied 
with $\alpha=2p=2m$ and $m=1,2,3\dots$,
the last phase factor is reduced to unity, 
and the ensuing
$\delta$-functions (and/or their derivatives) enforce the polygonal constraint \eqref{PC}. 

We need to stress that $\alpha$ being even is a sufficient  but not  necessary condition for the integral to vanish, if the polygonal constraint is not satisfied.  It is just that  for $\alpha$ even, the $\delta$-functions appear explicitly.
However, the charge neutrality condition \eqref{CNC} may be satisfied even for irrational and complex $\nu_n$ and $\alpha$, when the integral cannot be expressed in terms of $\delta$-functions. Recall from Eq.~\eqref{alpha1} that when the charge neutrality condition \eqref{CNC} is satisfied while the polygonal constraint \eqref{PC} is not, the integral vanishes due to zeros in the following function  
\begin{flalign}
\begin{split}
\frac{1}{\Gamma(\nu_M-\mu/2+1)}=\frac{1}{\Gamma(-m+1)};\ \ m = 1,2,3,\dots
\end{split}
\end{flalign}

\section{Conclusions} 
Integrals of products of Bessel functions have been studied for well over 150 years, and yet their properties and emergence in a variety of seemingly unrelated physical problems do not cease to amaze. Of particular interest is how the vanishing of such integrals 
are linked to two constraints, the polygonal constraint relating the argument coefficients to sides of a polygon, Eq.~\eqref{PC}, and what we have dubbed as the charge neutrality condition, Eq.~\eqref{CNC}. Given that the integral is convergent, 
if Eq.~\eqref{CNC} is satisfied 
while Eq.~\eqref{PC} is not,
the integral is identically equal to zero.

Examples of such integrals appear in a context of great interest in condensed matter physics, namely the density of states and mutual scattering rate of electrons moving on a 2D or 3D lattice.
The integrals that appear in this context involve Bessel functions of 
integer or half-integer orders, and $\delta$-functions may be employed to show intuitively why the integrals vanish when satisfying the constraints. For the model discussed, the absence of an umklapp scattering channel on a Fermi surface smaller than a critical value corresponds to the breaking of the polygonal constraint. The integral that gives the corresponding scattering rate goes from a finite value to exactly zero, with a concomitant discontinuity in the  derivative.
We also showed that the approach based on $\delta$-functions allows one to prove the vanishing of integrals of products of Bessel functions of arbitrary integer orders and in even dimensionalities $\alpha$, as defined in Eq.~\eqref{alpha}.
We expect these types of integrals to continue to appear in new and interesting settings, and as they emerge, further insights are likely to emerge alongside.

\section{Supplementary Material}
In the 
Supplementary Material, we present a detailed derivation of the umklapp scattering rate on a 2D honeycomb lattice. We begin in Sec.~I with a brief review of single-electron dynamics on a 2D honeycomb lattice in the nearest-neighbor-hopping approximation. We proceed in Sec.~IIA to define the on-site Hubbard Hamiltonian for electron-electron interactions in terms of the low energy electron operators presented in Sec.~I. Via the Boltzmann formalism and Fermi's Golden Rule, we derive 
the 
umklapp scattering rate in Sec.~IIB.
In Sec.~III we show how additional discontinuities in the derivative of $f(c)$ appear for a specific example. 

\acknowledgements
We thank A. Chubukov, I. Gaiur, T. Kiliptari, S. Majumdar, V. Roubtsov, V. Schechtman, C. Vignat, and V. Yudson for stimulating discussions. This work was initiated at the Aspen Center for Physics, which is supported by National Science Foundation grant PHY-2210452, and supported by National Science Foundation via grant DMR-2224000.   D. L. M. also acknowledges the hospitality of the Kavli Institute for Theoretical Physics, Santa Barbara, supported by the NSF grants PHY-1748958 and PHY-2309135.

\bibliography{jcoveyMend,dm_references}

\end{document}


\title{Integrals of Products of Bessel Functions:\\ 
An Insight from the Physics of Bloch Electrons\\
Supplementary Material}
\author{Joshua Covey and Dmitrii L. Maslov}
\affiliation{Department of Physics, University of Florida, Gainesville, FL 32611-8440, USA}
\email{jcovey@ufl.edu}
\date{\today}

\maketitle 
\tableofcontents

\section{\label{sec:basics}Basic Properties of Electrons on a \\ Two-Dimensional Honeycomb Lattice}
\begin{figure}[b]
  \includegraphics[scale=0.55]{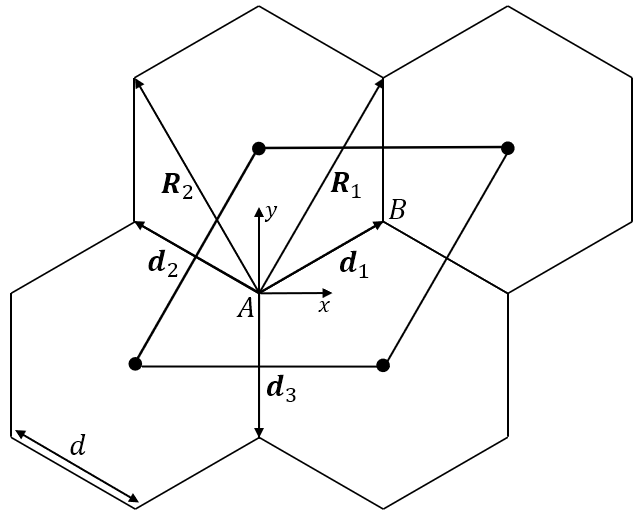}
  \caption{
  A unit cell of the honeycomb lattice may be chosen as any linear combination of the lattice vectors $\bR_1$ and $\bR_2$. A particular choice is a rhombus, containing two distinct lattice cites $A$ and $B$, separated by lattice constant $d$. With only nearest neighbor hopping allowed, particles on an $A$ ($B$) site can hop only to one of its three nearest neighbor $B$ ($A$) sites.}
  \label{fig:ab site}
\end{figure}

A unit cell of the real-space two-dimensional honeycomb lattice
can be chosen as 
a rhombus as in Fig.~\ref{fig:ab site}, which 
encloses two lattice sites, 
$A$ and $B$,
separated by 
lattice constant $d$. 
In the nearest-neighbor-hopping approximation, electrons are allowed to 
tunnel from an $A$ ($B$) site only to one of its three nearest-neighbor $B$ ($A$) sites.

The tight-binding Hamiltonian is given by  
\bea
H_0=-t\sum_{<i,j>}\bigl(a_{i}^\dagger b_{j} + h.c.\bigr),
\eea
where $a_i^\dagger$ ($a_i$) and $b_j^\dagger$ ($b_j$) are the operators creating (annihilating) electrons on the $i^\mathrm{th}$ $A$ site  and $j^\mathrm{th}$ $B$ site, respectively. (While we will account for spin later in the interaction term of the Hamiltonian, here we omit the spin index $s$ for brevity.) If the $i^\mathrm{th}$ A site is located at $\bR_i$, then the $j^\mathrm{th}$ $B$ site is located 
at $\bR_j=\bR_{in}=\bR_i+\bd_n$ with $n=1\dots 3$, where vectors $d_{1\dots 3}$ are shown in Fig.~\ref{fig:ab site}.
Alternatively, one can consider $a_i$ and $b_{i}$ with $n=1$ to correspond to electrons on the same lattice site,
such that the vectors connecting $\bR_i$ to its nearest neighbors are defined as $\bd_n=\{0,\bR_1,\bR_2\}$, where
\bea
\bR_1=\frac{d}{2}(\sqrt{3},3) \;\mathrm{and}\; \bR_2=\frac{d}{2}(-\sqrt{3},3).
\eea
The Fourier transforms (FTs) of the lattice operators are given by
\begin{flalign}
\begin{split}
a_{i}&=\frac{1}{\sqrt{N}}\sum_{\bk}e^{-i\bk\cdot\bR_i}a_{\bk} 
\;\mathrm{and}\;
b_{j}=b_{in}=\frac{1}{\sqrt{N}}\sum_{\bk}e^{-i\bk\cdot(\bR_i+\bd_n)}b_{\bk}, 
\end{split}
\end{flalign}
where $N$ is the total  number of unit cells.
On replacing all operators with their FTs, we have
\begin{flalign}
\begin{split}
H_0= -\frac{t}{N}\sum_{\mathclap{\substack{i,n,\\\bk,\bk'}}}\bigl(e^{i(\bk-\bk')\cdot\bR_i}e^{-i\bk'\cdot\bd_n}a_{\bk}^\dagger b_{\bk'} \ + \ h.c.\bigr).  
\label{ft h0}
\end{split}
\end{flalign}
With the help of an identity
\begin{flalign}
\begin{split}
\frac{1}{N}\sum_i e^{\pm i(\bk-\bk')\cdot\bR_i}=
    \delta_{\bk,\bk'},
\end{split}
\end{flalign}
we obtain
\begin{flalign}
\begin{split}
H_0= \sum_{\bk}(\varphi_\bk a_\bk^\dagger b_\bk + h.c.),\label{H0}
\end{split}
\end{flalign}
where 
\bea
\varphi_\bk&=&-t\sum_{n}e^{-i\bk\cdot\bd_n}. 
\eea
The eigenvalues of Eq.~\eqref{H0} are given by
\bea
\varepsilon_\bk &=& \pm|\varphi_\bk|  
=\pm t\sqrt{1+4\cos\frac{\sqrt{3}}{2}k_x d\left(\cos{\frac{\sqrt{3}}{2}k_xd}+\cos{\frac{3}{2}k_yd}\right)}.
\eea
Upon expanding to linear order about the $\bK$ point (see Fig.\ 2 of  the Main Text) so that $\bk\rightarrow\bK+\bk$ with $|\bk|\ll |\bK|$, the dispersion is reduced to a Dirac-like form
\bea
\varepsilon_\bk = \pm \vf |\bk|,
\eea
where the Fermi velocity is $\vf=3dt/2$ (with $\hslash=1$).
Correspondingly,
\bea
\varphi_\bk = \vf(k_x-ik_y). 
\eea
Expanding about $\bK'$ leads to the same result for the eigenvalues, while $k_x$ in $\varphi_\bk$ changes its sign, i.e. an electron in the $\bK'$ valley is the time-reversed copy of that in the $\bK$ valley. The low-energy Hamiltonian describing both the $\bK$ and $\bK'$ valleys can be written compactly as
\bea
H_0=\vf \sum_{\bk,\tau,s} (\tau k e^{-i\tau\phi_\bk}a_{\bk,\tau,s}^\dagger b_{\bk,\tau,s} + h.c.),
\eea
where $\tau=\pm1$ labels the  $\bK$ and $\bK'$ points, respectively, and
where the spin index $s=\ \up \text{or} \down$ is now displayed explicitly.

In general, $H_0$ may be diagonalized by the unitary transformation
\bea
a_{\bk,\tau,s}=\frac{1}{\sqrt{2}}(c_{\bk,\tau,s}+\varv_{\bk,\tau,s})
\;\mathrm{and}\;
b_{\bk,\tau,s}=\frac{\tau e^{i\tau\phi_\bk}}{\sqrt{2}}(c_{\bk,\tau,s}-\varv_{\bk,\tau,s}),
\eea
where $c^\dagger_{\bk,\tau,s}$ ($c_{\bk,\tau,s}$) and $\varv^\dagger_{\bk,\tau,s}$ ($\varv_{\bk,\tau,s}$) create (annihilate) particles in the upper and lower Dirac cones (conduction and valence bands),  respectively.  
However, if the Fermi energy ($\Ef$) is far above the charge-neutrality point, the valence band may be projected out so that
\bea
a_{\bk,\tau,s}=\frac{1}{\sqrt{2}}c_{\bk,\tau,s} 
\;\mathrm{and}\;
b_{\bk,\tau,s}=\frac{\tau e^{i\tau\phi_\bk}}{\sqrt{2}}c_{\bk,\tau,s}.
\label{ch of basis}
\eea
\section{Electron-Electron Scattering on a \\ Two-dimensional Honeycomb Lattice}
 \label{supp ref: deriv hc lattice}
\subsection{Interaction Hamiltonian and Scattering Probability}
We describe the electron-electron interaction via the on-site Hubbard model
\bea
H_U=\frac{U}{2}\sum_{i,s} : n_{i,s}n_{i,\bar{s}} :=\frac{1}{2}U\sum_{i,s} : \left(a^\dagger_{i,s} a_{i,s}+b^\dagger_{i,s} b_{i,s}\right)\left(a^\dagger_{i,\bar{s}} a_{i,\bar{s}}+b^\dagger_{i,\bar{s}} b_{i,\bar{s}}\right) :,
\eea
where $n_i$ is the number density operator on site $i$,
and if $s=\up$ then $\bar{s}=\down$, and vice versa. 
The normal ordering, denoted by $:A:$, accounts for subtracting off the zero-point energy.
In the momentum space, $H_U$ reads
\begin{flalign}
\begin{split}
H_U=\frac{U}{2N}\sum_{\mathclap{\substack{\bk,\bp,\\\bk',\bp',s}}} \delta_{\bk+\bp-\bk'-\bp',\bG} \Bigl(a^\dagger_{\bk',s}a^\dagger_{\bp',\bar{s}} a_{\bp,\bar{s}} a_{\bk,s} \ + \ b^\dagger_{\bk',s}b^\dagger_{\bp',\bar{s}} b_{\bp,\bar{s}} b_{\bk,s} \ + \ a^\dagger_{\bk',s}b^\dagger_{\bp',\bar{s}}b _{\bp,\bar{s}} a_{\bk,s} \ + \ b^\dagger_{\bk',s}a^\dagger_{\bp',\bar{s}}a _{\bp,\bar{s}} b_{\bk,s} \Bigr).
\end{split}
\end{flalign}
Note the total quasimomentum must be conserved up to an arbitrary reciprocal lattice vector $\bG$. 

We now go to the low energy regime, when the $a$ and $b$ operators can be replaced by Eq.~\eqref{ch of basis}. We consider only those intervalley scattering events in which the initial momenta $\bk$ and $\bp$ belong to the $\bK$ valley ($\tau=+1$), while  the final momenta $\bk'$ and $\bp'$ belong to the $\bK'$  valley ($\tau={-1}$), and vice versa. These events correspond to the largest change of the total momentum, and, therefore,  are the only ones which qualify as umklapp scattering. Upon keeping only such terms, $H_U$ is reduced to
\begin{flalign}
\begin{split}
H_U=\frac{U}{8N}\sum_{\mathclap{\substack{\bk,\bp,\\\bk',\bp',\tau,s}}} \delta_{\bk+\bp-\bk'-\bp',\Delta\bK} \Bigl(1-e^{i\tau(\phi_\bk+\phi_{\bk'})}\Bigr) \Bigl(1-e^{i\tau(\phi_\bp+\phi_{\bp'})}\Bigr) c^\dagger_{\bk',-\tau,s}c^\dagger_{\bp',-\tau,\bar{s}}c_{\bp,\tau,\bar{s}}c_{\bk,\tau,s}.
\end{split}
\end{flalign}
\subsection{Umklapp Scattering Rate}
The time evolution of a distribution function for a spatially homogeneous system with  electron-electron collisions is described by the Boltzmann equation
\begin{flalign}
    \begin{split}
        \frac{\partial f_{\bk}}{\partial t}= -I_{\mathrm{ee}}[f_{\bk}],
    \end{split}
\end{flalign}
where the collision integral is given by
\begin{flalign}
    \begin{split}
        I_{\mathrm{ee}}[f_{\bk}]=\sum_{\mathclap{\substack{\bp,\bk',\bp',\\\tau,s,s'}}}\widetilde W_{\bk,\bp;\bk',\bp';\tau,s,s'}\ [f_\bk f_\bp(1-f_{\bk'})(1-f_{\bp'}) - f_{\bk'}f_{\bp'}(1-f_\bk)(1-f_\bp)].
        \label{coll int}
    \end{split}
\end{flalign}
The scattering rate can be found via Fermi's golden rule as
\begin{flalign}
    \begin{split}
\widetilde W_{\bk,\bp;\bk',\bp';\tau,s,s'}&=2\pi\bigl|\bigl\langle\bk',-\tau,s';\bp',-\tau,\bar{s}'\bigl|H_U\bigr|\bk,\tau,s;\bp,\tau,\bar{s}\bigr\rangle\bigr|^2 \delta(\varepsilon_\bk+\varepsilon_\bp-\varepsilon_{\bk'}-\varepsilon_{\bp'}),
        \label{mat elements}
    \end{split}
\end{flalign}
where $\ve_\bk$ is the electron energy measured relative to the Fermi energy,  
and the wavefunctions are obtained by the following operators acting on the vacuum state $\bigl|0\bigl\rangle$
\begin{flalign}
    \begin{split} \bigr|\bk,\tau,\up;\bp,\tau,\down\bigr\rangle=c^\dagger_{\bk,\tau,\up}c^\dagger_{\bp,\tau,\down}\bigl|0\bigl\rangle.
    \end{split}
\end{flalign}
As stated previously, the wavefunctions are chosen such that the two electrons start in the same valley and end up in the opposite valley. Then, 
\bea
        \widetilde W_{\bk,\bp;\bk',\bp';\tau,s,s'}&=& 2\pi\frac{U^2}{N^2} \left[ \sin^2\left(\frac{\phi_\bk+\phi_{\bk'}}{2}\right)\sin^2\left(\frac{\phi_\bp+\phi_{\bp'}}{2}\right)\delta_{s=s'} + \sin^2\left(\frac{\phi_\bk+\phi_{\bp'}}{2}\right)\sin^2\left(\frac{\phi_\bp+\phi_{\bk'}}{2}\right)\delta_{s=\bar{s}'}\right]\nn\\
        &&\times \delta_{\bk+\bp-\bk'-\bp',\Delta \bK}\ \delta(\varepsilon_\bk+\varepsilon_\bp-\varepsilon_{\bk'}-\varepsilon_{\bp'}).  
        \label{mat elements 2}
        \eea
Substituting this result into Eq.~\eqref{coll int}, we obtain
\begin{flalign}
    \begin{split}
        I_{\mathrm{ee}}[f_{\bk}]&=\frac{16\pi U^2}{N^2}\sum_{\mathclap{\substack{\bp,\bk',\bp'}}} 
        \sin^2\left(\frac{\phi_\bk+\phi_{\bk'}}{2}\right)\sin^2\left(\frac{\phi_\bp+\phi_{\bp'}}{2}\right)  [f_\bk f_\bp(1-f_{\bk'})(1-f_{\bp'}) - f_{\bk'}f_{\bp'}(1-f_\bk)(1-f_\bp)]
        \\ 
        &\hspace{8cm}\times \ \delta_{\bk+\bp-\bk'-\bp',\Delta \bK}\ \delta(\varepsilon_\bk+\varepsilon_\bp-\varepsilon_{\bk'}-\varepsilon_{\bp'}).
        \label{coll int 2}
    \end{split}
\end{flalign}

Near equilibrium, the distribution function may be written as  
\bea
f_\bk=n_\bk+\delta f_\bk,
\eea
where 
\bea
n_\bk=\frac{1}{e^{\ve_\bk/T}+1} \hspace{.5cm} (k_B=1)
\eea
is the Fermi function and $|\delta f_\bk|\ll n_\bk$.
The scattering rate of an electron with momentum $\bk$ is found by taking a functional derivative of the collision integral with respect to a small variation of the distribution function \cite{levinson:book}
\bea
    \frac{1}{\tau_\bk}&=&-\frac{\delta}{\delta f_{\bk}}\frac{\partial f_\bk}{\partial t}=\frac{\delta}{\delta f_{\bk}}I_{\mathrm{ee}}[f_{\bk}].
\eea
When distributing the functional derivative to all terms in the integrand of $I_{\mathrm{ee}}$, we note that one of the sums over the momenta  will collapse if $\delta/\delta f_\bk$ acts on any distribution function with momentum different from $\bk$ due to $\delta f_{\bk'}/\delta f_\bk=\delta_{\bk,\bk'}$. Now we replace the sums over the two remaining momenta by the integrals as $\sum_\bk=A\int d^2k/(2\pi)^2$ with $A$ being the area of the system, define a coupling constant $u=UA/N=UA_0$ with $A_0$ being the area of a unit cell, and  introduce  a continuous $\delta$-function as $ \delta(\bk+\bp-\bk'-\bp'-\Delta \bK)=\lim_{A\to\infty}A\ \delta_{\bk+\bp-\bk'-\bp',\Delta \bK}$.  With these definitions, we see that the corresponding term in $1/\tau_\bk$ behaves as $1/A$ and thus can be discarded at $A\to\infty$. On the other hand, when $\delta/\delta f_\bk$ acts on $f_\bk$, there are no factors of $1/A$ left.
Therefore, $1/\tau_\bk$ is 
given by 
\begin{flalign}
    \begin{split}
        \frac{1}{\tau_\bk}&=\int \frac{d^2p}{(2\pi)^2} \int \frac{d^2k'}{(2\pi)^2}\int \frac{d^2p'}{(2\pi)^2} W_{\bk,\bp;\bk',\bp'} \ [n_\bp(1-n_{\bk'})(1-n_{\bp'}) + n_{\bk'}n_{\bp'}(1-n_\bp)]
        \\ 
        &\hspace{8cm}\times \ \delta(\bk+\bp-\bk'-\bp'-\Delta \bK)\ \delta(\varepsilon_\bk+\varepsilon_\bp-\varepsilon_{\bk'}-\varepsilon_{\bp'}), 
        \label{coll int 2}
    \end{split}
\end{flalign}
where 
\begin{flalign}
    \begin{split} 
        W_{\bk,\bp;\bk',\bp'}=16\pi u^2 \sin^2\left(\frac{\phi_\bk+\phi_{\bk'}}{2}\right)\sin^2\left(\frac{\phi_\bp+\phi_{\bp'}}{2}\right), 
    \end{split}
    \end{flalign}
    which coincides with Eq.~(23) of the Main Text.
    
Since we are considering isotropic Fermi surfaces, the integral of the momentum can be written as $\int d^2p=\int_0^\infty dpp\int_0^{2\pi}d\phi_\bp$, and the same for other momenta.  To find the leading order contribution for $T\ll \Ef$, we need only consider those electrons close to the Fermi surface, so that $|\bk|=|\bp|=|\bk'|=|\bp'|=\kf$, $\ve_\bp=\vf(p-\kf)$, and $dpp=(\kf/\vf)d\ve_\bp$. Additionally, the $\delta$-function enforcing energy conservation may be rewritten as 
\bea
\delta(\varepsilon_\bk+\varepsilon_\bp-\varepsilon_{\bk'}-\varepsilon_{\bp'})=\int d\omega \ \delta(\ve_\bk-\ve_{\bk'}-\omega)\delta(\ve_\bp-\ve_{\bp'}+\omega).
\eea
Finally, it is convenient to average $1/\tau$ 
over the directions of $\bk$.  The average scattering rate is then given by
\begin{flalign}
\begin{split}
\left\langle\frac{1}{\tau}\right\rangle &= \frac{4}{\pi}\frac{\kf^3u^2}{\vf^3}\int_{-\infty}^{\infty} d\omega  \int_{-\infty}^{\infty} d\ve_\bp \frac{1}{1-n(\ve_\bk)}n(\ve_\bp)\Bigl[1-n(\ve_\bk-\omega)\Bigr]\Bigl[1-n(\ve_\bp+\omega)\Bigr] \\
&\times \ \frac{1}{(2\pi)^5} \int_0^{2\pi} d\phi_{\bk} \int_0^{2\pi} d\phi_{\bp} \int_0^{2\pi} d\phi_{\bk'} \int_0^{2\pi} d\phi_{\bp'} \sin^2\left(\frac{\phi_\bk+\phi_{\bk'}}{2}\right)\sin^2\left(\frac{\phi_\bp+\phi_{\bp'}}{2}\right)\\
&\times\delta(\bk+\bp-\bk'-\bp'-\Delta \bK),
\end{split}
\end{flalign}
where we used the equality 
\bea
n_\bk n_\bp(1-n_{\bk'})(1-n_{\bp'}) = n_{\bk'}n_{\bp'}(1-n_\bk)(1-n_\bp).
\eea 
The integrals over $\omega$ and $\ve_\bp$ give
\begin{flalign}
\begin{split}
 \int_{-\infty}^{\infty} d\omega  \int_{-\infty}^{\infty} d\ve_\bp \frac{1}{1-n(\ve_\bk)}n(\ve_\bp)\Bigl[1-n(\ve_\bk-\omega)\Bigr]\Bigl[1-n(\ve_\bp+\omega)\Bigr] = \frac{1}{2}\Bigl(\pi^2T^2+\ve_\bk^2\Bigr). 
\end{split}
\end{flalign}
For an electron right on the Fermi surface ($\ve_\bk=0$), we obtain
\begin{flalign}
\begin{split}
\left\langle\frac{1}{\tau}\right\rangle &= \frac{\pi}{2} \left(\frac{\kf u}{\vf}\right)^2 \frac{T^2}{\Ef} \frac{1}{(2\pi)^5} \int_0^{2\pi} d\phi_{\bk} \int_0^{2\pi} d\phi_{\bp} \int_0^{2\pi} d\phi_{\bk'} \int_0^{2\pi} d\phi_{\bp'} 
\Bigl[1-\cos(\phi_\bk+\phi_{\bk'})\Bigr] 
\Bigl[1-\cos(\phi_\bp+\phi_{\bp'})\Bigr]  \\
& \times \ \delta(\cos\phi_\bk + \cos\phi_\bp - \cos\phi_{\bk'} - \cos\phi_{\bp'} - c\cos\phi_{\Delta\bK})\delta(\sin\phi_\bk + \sin\phi_\bp - \sin\phi_{\bk'} - \sin\phi_{\bp'} - c\sin\phi_{\Delta\bK})\\
&=\frac{\pi}{2} \left(\frac{\kf u}{\vf}\right)^2 \frac{T^2}{\Ef} f\left(\frac{|\Delta \bK|}{\kf}\right),
\end{split}
\end{flalign}
which is Eq.~(24) of the Main Text.

\section{Discontinuous Derivatives}
 An integral over a product of Bessel functions is repeated here in its general form
\begin{equation}
    \begin{split}
   f(\{c_n\})=
\int^\infty_0 d\rho \rho^{\alpha-1} \prod_{n=1}^{M>1}J_{\nu_n}(c_n \rho). \label{alpha} 
    \end{split}
\end{equation}
In addition to the discontinuity at the threshold for the polygonal constraint to be satisfied [Eq.~(2) of the Main Text], such an integral  
may have additional discontinuities in its derivatives with respect to $c_1\dots c_M$ well below the threshold. An example of such behavior is shown in Fig.~4 of the Main Text. These additional discontinuities may be deduced by the large-argument behavior of the Bessel functions.
Recalling the asymptotic form of a Bessel function for large argument, the integrand may be written
\beq
\rho^{\alpha-1}\prod^{M>1}_{i=1} J_{\nu_n}(c_n \rho)\bigg|_{\rho\rightarrow\infty} \rightarrow 
g(\rho, \{c_n\})=\rho^{\alpha-1}\prod^{M>1}_{i=1} \sqrt{\frac{2}{\pi c_n \rho}}\cos\Bigl(c_n\rho-\frac{\pi}{2}\nu_n-\frac{\pi}{4}\Bigr).
\eeq
Consider the case where all $\nu_n=0$ and all $c_n=1$ except for $c_M=c$. Taking a derivative of the product with respect to $c$ gives
\beq
\frac{\partial g}{\partial c}\propto \rho^{\alpha-\frac{M}{2}} \cos^{M-1}\Bigl(\rho-\frac{\pi}{4}\Bigr)\sin\Bigl(c\rho-\frac{\pi}{4}\Bigr),
\eeq
where this extra factor of $\rho$ changes the integral's criterion for convergence. We take $S_1$ from the Main Text as an example, where $\alpha=2$ and $M=5$, so that 
\beq
\frac{\partial g}{\partial c}\propto \frac{1}{\sqrt{\rho}} \cos^4\Bigl(\rho-\frac{\pi}{4}\Bigr)\sin\Bigl(c\rho-\frac{\pi}{4}\Bigr).
\label{partial g}
\eeq
Using basic trigonometric identities, it is easy to show that Eq.~\eqref{partial g} is a sum of the following three terms
\beq
\frac{\partial g}{\partial c}\propto \frac{1}{\sqrt{\rho}} \left[ \frac{3}{8} + 
\frac 12\cos\left(2\rho-\frac{\pi}{2}\right) + 
\frac 18\cos\left(4\rho-\pi\right)\right] \sin\Bigl(c\rho-\frac{\pi}{4}\Bigr).\label{sum} 
\eeq
The first term becomes non-oscillatory when $c=0$, 
and thus the integral for $\partial f(c)/\partial c$ diverges at the upper limit.
Upon applying yet another trigonometric identity,   
we see the second and third term in Eq.~\eqref{sum} become non-oscillatory at $c=2$ and $c=4$, respectively, which implies the discontinuities of $\partial f(c)/\partial c$ at these values of $c$, in addition to $c=0$. For the example considered, $c=4$ corresponds to the threshold for the polygonal constraint to be satisfied.

\bibliography{jcoveyMend,dm_references}